# Coarse grained modeling of a metal-organic framework/polymer composite and its gas adsorption at the nanoparticle level

*Cecilia M. S. Alvares[a], Rocio Semino[b*]*

[a] ICGM, Univ. Montpellier, CNRS, ENSCM, Montpellier, France

[b] Sorbonne Université, CNRS, Physico-chimie des Electrolytes et Nanosystèmes Interfaciaux, PHENIX, F-75005 Paris, France

## Abstract

Simulations have acted as a cornerstone to understand MOF/polymer interface structure, however, no molecular-level simulation has yet been performed at the nanoparticle scale. In this work, a hybrid MARTINI/Force Matching (FM) force field was developed and successfully implemented to model the ZIF-8/PVDF composite at a coarse grained resolution. Inter-phase interactions were modeled using FM potentials, which strive to reasonably reproduce the forces from an atomistic benchmark model, while intraphase interactions are modeled using the general-purpose MARTINI potentials. Systems made of a ZIF-8 nanoparticle embedded into a PVDF matrix were considered to evaluate the effect of nanoparticle size and morphology in the polymer structure and in the CO2 adsorption. Results show that simulations at the nanoparticle level are crucial for depicting the polymer penetration. Notably, the smallest nanoparticle exhibited the least extent of polymer penetration, while the cubic nanoparticle exhibited the highest amount. Polymer conformation and local density values change similarly in all ZIF-8/PVDF systems depending on whether the polymer lies inside or outside of the nanoparticle domain. All composite models present more significant $CO_2$ adsorption in the nanoparticle domain than in the PVDF phase, in agreement with experiments. More remarkably, the small rhombic dodecahedron ZIF-8/PVDF system presents a larger equilibrium amount of gas adsorbed at ambient condition compared to the other two systems, in alignment with the observed polymer penetration trend. On the other hand, the amount of $CO_2$ adsorbed at equilibrium is lower for the rhombic dodecahedron morphology than for the cubic one, contrary to the intuitive expectation founded in the polymer penetration trend. This result could be a reflection of a difference in the number of surface adsorption sites.

**Introduction**

Mixed matrix membranes formed by metal-organic frameworks (MOFs) dispersed into a polymer matrix provide an attractive alternative solution to current pure-polymer membrane-based technologies. Indeed, these composite materials ally the excellent merit parameters of MOFs for adsorption-related applications[1] with the ease of processability of polymers.[2,3] One of the key aspects of the optimization of MOF/polymer MMMs for target applications is to engineer their interfaces. For example, numerous strategies to prevent the formation of defects have been developed,[4] including nanoparticle coating strategies,[5,6] modifications or functionalizations to enhance chemical compatibility,[7–10] cross linking of MOF/polymer composites,[11,12] and controlled fusion of MOF and polymer components,[13] to cite a few examples. Since defects can contribute favorably to gas separation performances as well,[14] deliberate defect introduction has also been explored.[15,16]

It is nowadays impossible to obtain a molecular level picture of the MOF/polymer interface, let alone the interfacial defects, from a direct experiments standpoint. For this reason, simulations have played a fundamental role in this field. Semino and coworkers developed the first microscopic methodology to model MOF/polymer interfaces that allows to obtain detailed structural information that is consistent with experimental data.[17] This methodology was further applied to study a series of MOF-polymer pairs,[18] allowing to describe compatibility (defined as the ability to interact (strongly or poorly)) at the molecular level for the first time. It was found that systems that were 'poorly' compatible, had interfacial microvoids as a result of the MOF/polymer repulsion.[17] On the contrary, systems that have 'good' compatibility, exhibit high MOF/polymer adhesion, resulting in a partial penetration of the polymer chains into the MOF porosity.[19] The non straightforward relationship between compatibility and performance for gas separation was also explored.[14] However, computational cost prevents using these models to study some crucial aspects such as the impact of the filler size and morphology on both the interface structure and gas adsorption properties. Instead, studying the MOF/polymer system at this scale requires coarser approaches.

A few works addressing modeling MOFs via coarse graining (CG) strategies have been published up to date. Dürholt and collaborators have developed the first particle-based CG model of a MOF, HKUST-1, based on a genetic algorithm minimization of the difference between the Hessian matrices of the CG and a reference all-atom (AA) benchmark.[20] Following this publication, Alvares and coworkers have performed a systematic study of particle-based CG models applied to MOFs.[21,22] These include general-purpose force fields such as MARTINI,[23] as well as *ad hoc* CG force field fitting approaches that strive to reproduce either reference forces (force matching (FM), also referred to as MS-CG method)[24] or structural properties (Iterative Boltzmann Inversion).[25] In particular, FM CG force fields for MOFs have been able to reproduce the swing effect,[22] a subtle phase transition that ZIF-8 undergoes upon gas loading.[26,27] Other innovative strategies for modeling MOFs at a coarser resolution have been recently proposed.[28,29] The first CG-level study of MOF/polymer composites shed light into the spatial extent of structural perturbations produced in the polymer due to the presence of the MOF.[30] However, it focused on a MOF slab surrounded by polymer only so no edges, curvature or vertices were taken into consideration. Moreover, even though the previously mentioned genetic algorithm optimization strategy was applied to adequately treat intraphase interactions, cross-interactions posed issues and had to be manually tuned to reproduce an adequate MOF/polymer degree of interaction.[30] In this sense, the previous success in employing FM to model cross gas/host interactions with such a degree of chemical specificity that even the swing effect was reproduced, suggests that this strategy could also be employed to model cross-interactions for MOF-polymer pairs.

In this work, the ZIF-8/polyvinylidene fluoride (PVDF) composite, which has been successfully applied for liquid and gas separations,[31,32] is studied at the mesoscale level (scales within $\approx$ 20 nm) to unveil the effect of the filler size and morphology in polymer structuration and gas adsorption. Indeed, these are non trivial aspects that would benefit from a deeper understanding.[33-35] A CG force field strategy that combines the general-purpose MARTINI 3 potentials to model intraphase interactions with the chemically specific FM potentials to model the interphase interactions is presented. The need for such a strategy is shown by presenting results for simulations made with a force field composed solely of MARTINI potentials, which predict an excessively attractive MOF/polymer interaction. The hybrid MARTINI/FM force field solves this problem, and is then used to study the composite system at the mesoscale level. Three different nanoparticles are considered as fillers. Notably, despite the polymer penetration into the MOF phase predicted by the hybrid MARTINI/FM force field, it is shown that $CO_2$ molecules are still adsorbed in greater amounts in the ZIF-8 nanoparticle domain than in the polymer phase outside the overlap region, in agreement with experiments. We believe that this hybrid MARTINI/FM CG strategy or in general, mixing generic and chemically-specific ad hoc approaches, is particularly promising for modeling MOF/polymer and other complex interfaces at the mesoscale.

This article is structured as follows. Section II discusses the CG force field development, creation of the simulation domain to study the ZIF-8/PVDF systems, and the methodology for ultimately validating the models. Molecular dynamics (MD) and Monte Carlo (MC) simulations, all performed in LAMMPS[36] are also discussed. Finally, the results and conclusions are summarized in Sections III and IV respectively. The work also features a Supporting Information (SI), in which further information and comments associated with the methodology, simulation details as well as further results can be found.

**Methodology**

The methodology is divided into sections for force field development and validation. Details on simulations used throughout the work are also given. Whenever MD simulations are performed, the Nosé Hoover equations of motion as originally developed either to sample the NVT or NPT ensemble were used for the dynamics.[37,38] These equations are correspondingly referred to along the text as "NVT equations of motion" or "NPT equations of motion". Unless specified differently, the damping constants for the thermostat and barostat (whenever applicable) are, respectively, 100x and 1000x the value of the timestep used for the numerical integration, in accordance with the recommendation in the LAMMPS manual.[39] The polydispersity of the polymer phase in all AA and CG simulations made within this work for the ZIF-8/PVDF systems as well as for bulk PVDF is the same and can be found shown in figure 1(c) as a function of the number of PVDF monomers ($-CH_2-CF_2-$).

*1 Development of the CG force field*

Studying ZIF-8/PVDF at the CG level requires defining a mapping and a force field. The mappings of choice for the MOF and polymer are shown in figures 1(a) and (b), respectively, by indicating the groups of atoms that are repeatedly replaced by a bead along the ZIF-8 and PVDF phases. All beads are assigned a bead type and beads that replace chemically identical groups of atoms share the same bead type. For ZIF-8, the beads replace either a $Zn^{2+}$ cation or a 2-methyl imidazole ligand (2mIm$^-$), leading to a total of two bead types. For PVDF, four bead types exist: beads type 1 replace two sequentially bonded monomers, $-CH_2-CF_2-CH_2-CF_2-$, beads type 2 and 3 correspond to terminal beads containing a carbon bonded to three hydrogen and fluor atoms ($CH_3-CF_2-$ and $CF_3-CH_2-$), respectively, and beads type 4 replace a single monomer ($-CH_2-CF_2-$). The latter only exist in chains formed by an odd number of monomers, and are defined to always be in-between beads type 1 and 3 of that same chain. The mass of all beads was assigned to be the sum of the masses of the atoms they replace.

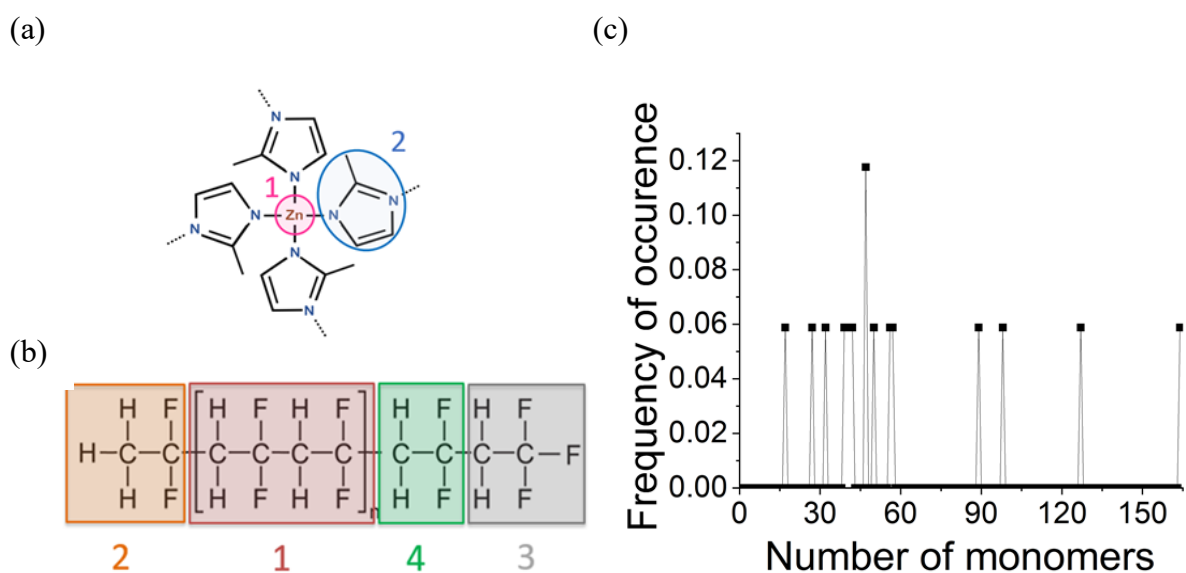

Figure 1. Representation of groups of atoms that are replaced by a single bead within the mappings chosen to study (a) ZIF-8 and (b) PVDF in CG resolution. Bead types are shown indexed. (c) Polydispersity of the polymer model (expressed in terms of the number of atomistic monomers (i.e., a $-CH_2-CF_2-$ group)).

The force field to study ZIF-8/PVDF is defined as a sum of bonded and pairwise non-bonded potentials. The former encapsulate bond and angle potentials that, as usual, are functions of bond length and 3-body angle values, respectively. The choice of a potential fitting strategy to derive the bonded and non-bonded potentials was made taking into consideration previous works in which different potential fitting strategies were investigated for bulk ZIF-8.[21,22] This is because finding suitable CG fitting strategies to model interactions for ZIF-8 is more challenging than for PVDF, since CG force fields stemming from different potential fitting strategies have already been deployed to study polymers, yielding, in particular, good structural results.[40-44] As a further criterion to pre-evaluate the suitability of the force field, the polymer penetration in the ZIF-8 phase was assessed at the CG level relying on a ZIF-8/PVDF slab system (see fig. S2) prior to studying the binary system at the nanoparticle level. Although the amount of polymer penetration is not known for this particular case, several experimental works report an increase in total free volume in ZIF-8/PVDF membranes compared to the pure PVDF counterpart,[45-48] suggesting thus that any possibly occurring polymer penetration should be modest.

Based on the simplicity of MARTINI force fields and their surprisingly good performance in reproducing the structure of ZIF-8 at highly coarsened resolutions such as the one chosen herein,[21] a force field composed of MARTINI potentials was initially considered. MARTINI 3 was chosen over MARTINI 2 due to the fact that it yields force fields better able to describe the elastic deformations of ZIF-8.[21] Following that choice, beads type 1 of PVDF are classified as RX1 while beads type 2, 3 and 4 are all classified as SX1 within MARTINI 3. This classification of bead flavors follows from the fact that PVDF is highly hydrophobic as well as strives for bead sizes better aligned with the recommendations while, simultaneously, allowing to better distinguish their relative sizes within the mapping of choice. Once the bead flavors for the polymer are defined, the parameters for the non-bonded potentials involving PVDF bead pairs are automatically given within MARTINI. For parameterizing the bonded potentials, MARTINI prescribes in its guidelines that the values of the parameters should be such that reference bond and angle distribution functions (BDFs and ADFs) are reproduced for the system in the desired resolution.[49] For this purpose, reference BDFs and ADFs for the bulk polymer in the CG resolution shown in fig. 1(b) were considered. Each BDF or ADF concerns a distinct sequence of two or three bead types which replace atoms that were chemically bonded in the AA representation. These distributions were built using in-house python codes which coarsen atomistic configurations attained in AA-MD simulations made at ambient temperature and pressure conditions for bulk PVDF. As prescribed within MARTINI 3, CG beads are centered in the center of geometry (COG) of the respective group of atoms they are associated with in the AA → CG mapping when coarsening the configurations. Specifics of the atomistic simulations and of the computation of the reference ADFs and BDFs can be found in section 1 of the SI.

Notably, the reference BDFs and ADFs for PVDF at the CG level, shown in figures S1(a) and S1(b) of the SI, revealed rather asymmetric profiles, in some cases featuring more than one peak. Such profiles are indeed expected since the relative positioning of the beads can change significantly depending on the diversity of conformations adopted by the chains. Yet, these cannot be reproduced using a force field where bonded potentials have the analytical form prescribed within MARTINI, shown in equations (1) and (2) for the bond and angle potentials, respectively:

$$U_{bond} = \frac{1}{2} k_b (r - r_o)^2 \tag{1}$$

$$U_{angle} = \frac{1}{2} k_\theta (cos(\theta) - cos(\theta_o))^2 \tag{2}$$

In these equations, r and θ denote instantaneous bond lengths and 3-body angle values, $k_b$ and $k_\theta$ are parameters referred to as force constants, and $r_o$ and $\theta_o$ are parameters playing the role of equilibrium bond length and angle values. As the case is more drastic in the ADFs profiles, angle potentials for the polymer were dismissed from the force field to avoid reinforcing an incorrect structure, while bond potentials were set over all sequences of two consecutive beads in the chains. A distinct bond potential was defined for each distinct pair of bead types. For each bond potential, the equilibrium bond length was determined by averaging the occurring instantaneous values reported in the respective bond distribution function. The force constants were iteratively optimized by doing CG-MD simulations at ambient temperature and pressure conditions of a bulk polymer using a MARTINI force field given by a sum of bond potentials and non-bonded potentials underlying the choice of bead flavors previously made. Once the MARTINI force field contemplating the optimal bond potentials for reproducing BDFs was reached, the corresponding density for the bulk CG PVDF at ambient (T,P) was computed for further comparing with values attained for the polymer in the CG ZIF-8/PVDF systems. Details of the CG-MD simulations used in the parametrization of these bond potentials, computation of the BDFs and of the average density for the bulk CG polymer can be found in section 2 of the SI.

Bonded and non-bonded potentials stemming from model mBM3III, investigated in previous works for bulk ZIF-8,[21] were borrowed to model the interactions between ZIF-8 beads in the ZIF-8/PVDF MARTINI force field. This model features bond and angle potentials as bonded contributions and bead flavors SP5q and SP2aq for beads type 1 and 2, respectively, which, put together with the classification of bead flavors for PVDF, defines also the non-bonded interactions between ZIF-8 and PVDF beads. Notably, this ZIF-8 model was chosen because its underlying bead flavors should lead to the highest degree of separation between the ZIF-8 and PVDF phases within the MARTINI miscibility table[23] amongst all the other MARTINI force fields investigated in previous works.[21] This is expected to better match the increase in free volume observed experimentally upon loading a PVDF membrane with ZIF-8 fillers.[45-48] Finally, when assembling the force field for the binary system, no differentiation in bead types was made for ZIF-8 beads at (nor near) the surface of the ZIF-8 phase to account for the presence of undercoordinated $Zn^{2+}$ and 2mIm$^-$ ligands or possible terminal groups bonded to them.

In order to test the quality of the assembled ZIF-8/PVDF MARTINI force field, a simulation of a CG ZIF-8/PVDF slab system was carried out to assess the extent of polymer penetration. An initial configuration was created by coarsening an all-atom configuration of a ZIF-8 slab borrowed from previous works[17] (shown in fig. S2) and putting a total of 68 coarsened PVDF chains along the direction normal to the slab surface. The AA slab originally corresponds to a ZIF-8 supercell cut in the [011] plane and terminated with -OH and -H groups, whose presence was ignored during the coarsening. No initial overlap between the two phases was introduced when creating the configuration, rather, void existed between them as well as between some of the polymer chains. Subsequently, an MD-based approach was deployed to reach a more realistic configuration for the ZIF-8/PVDF CG slab system (i.e., one that does not contemplate the artificially created void). Details can be found in section 3 of the SI. From this configuration, a CG-MD simulation was made using the NPT equations of motion with a timestep of 20 fs, as typically prescribed for MARTINI force fields.[23,50] The target temperature and pressure were set to 300 K and 1 atm respectively. Configurations for the equilibrated system were collected in order to build single-configuration bead density profiles along the direction normal to the ZIF-8 surface using an in-house python code aiming to evaluate the evolution of polymer penetration. Further details of this calculation can also be found in section 3 of the SI.

The density profiles for the ZIF-8 (light blue) and PVDF (dark red) beads for a configuration attained after a total of 6M timesteps are shown in fig. 2(a), and accuse a very discouraging amount of polymer penetration. The polymer extends throughout the entire length of the ZIF-8 phase and has local density values that peak to values as high as the ones observed outside the overlap region, in which PVDF density values are similar to the ones obtained for the bulk CG system. This can be associated with a significant filling of the MOF pores. While the real extent of PVDF penetration in ZIF-8 is not known, this result does not seem aligned with the experimental constatation of an increase in free volume upon incorporating ZIF-8 nanoparticles into PVDF.

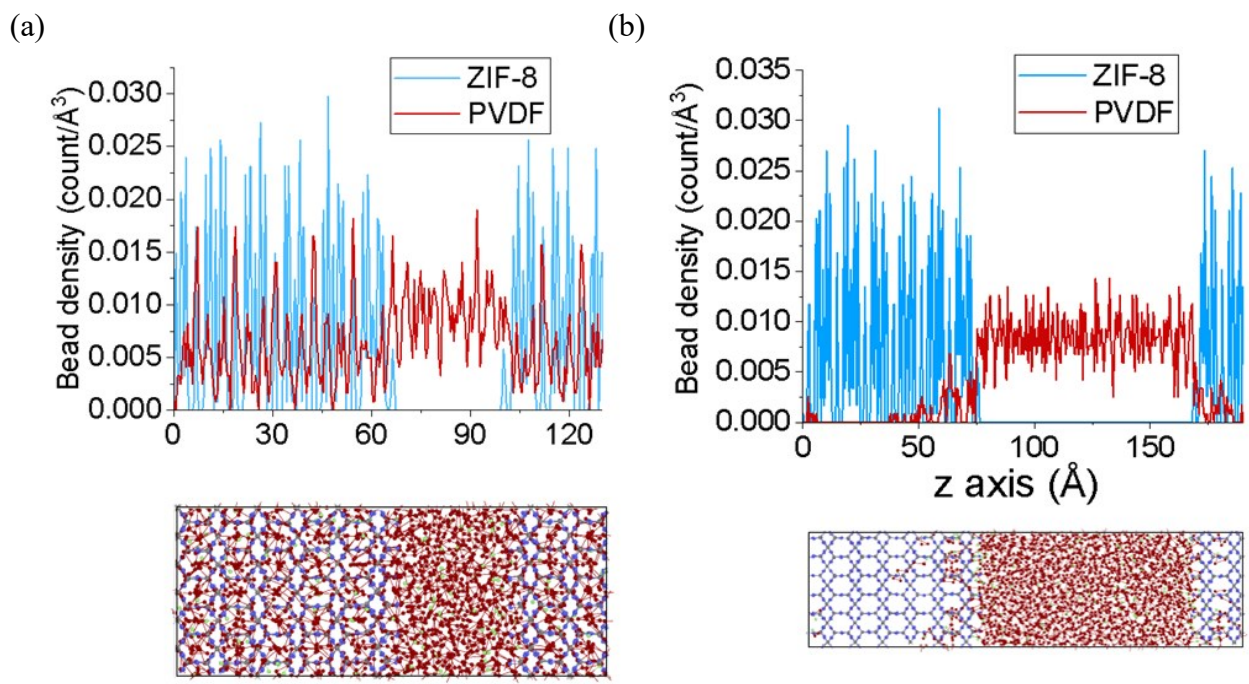

Figure 2. ZIF-8 (light blue) and PVDF (dark red) bead density profiles for a ZIF-8/PVDF slab system modeled at the CG level obtained with (a) the full-MARTINI force field first developed and (b) the hybrid MARTINI/FM force field. For further reference, the snapshot of the configuration from which the profiles are built is also shown (gray and blue beads: ZIF-8 beads types 1 and 2; dark red beads: PVDF beads types 1 and 4; light green beads: PVDF beads types 2 and 3). Aiming to keep the same size for the plots in figure (a) and (b) as well as have a correspondence between the x-axis in the plots and the z direction of the ZIF-8/PVDF slab system, the configuration in figure 2(a) features a larger cross section than the one associated with figure 2(b) as a consequence of the smaller size in the z direction due to a more extensive polymer penetration.

Aiming to improve the results, the force field was revisited. Potentials concerning beads from a same phase were kept unchanged, but FM non-bonded potentials were used for the

MOF/polymer cross interactions instead. Given that a FM force field strives to reproduce reference values of forces experienced by the beads in a set of reference CG configurations, the idea is that using cross potentials coming from FM should allow for a more rightful force contribution coming from the interactions between beads of the two phases. Thus, naturally, the polymer penetration may be more adequately depicted by this hybrid force field. A brief overview of the FM algorithm is discussed in the SI (section 5). It may be worth noting that CG force fields derived by combining different potential fitting strategies have already been proposed in the literature.[51]

For the sake of fitting FM potentials, 4500 configurations coming from AA-MD simulations of the ZIF-8/PVDF slab system were used. The simulations concern the AA version of the ZIF-8 slab and PVDF chains used in the CG-MD simulation, only that the number of PVDF chains is larger in the CG case. The atomistic configurations concern a thermodynamic state having (T = 300 K, p = 1 atm) and were sampled in the NPT ensemble. Details of these simulations can be found in the SI (section 4). The AA configurations of the ZIF-8/PVDF slab system to be used within the FM algorithm were coarsened centering the beads at the COG of the respective group of atoms, in alignment with what is prescribed within MARTINI 3. The reference force experienced by each bead in the CG resolution was defined as the sum of the forces experienced by the atoms linked to it within the AA → CG mapping. In that case, only the forces stemming from the non-bonded potentials composing the AA force field were considered.

The CG force field was defined within the algorithm as a sum of only pairwise non-bonded potentials between (ZIF-8)-(ZIF-8), (ZIF-8)-(PVDF) and (PVDF)-(PVDF) beads. These potentials were given by cubic splines of bin size 0.005 Å and having a cutoff of 15 Å. One non-bonded potential for each distinct type of bead type pair was set and the decision of not distinguishing between surface and bulk ZIF-8 beads was kept. Notably, the approach previously set for defining the CG force field and calculating the reference forces implies that all the non-bonded potentials defined in the CG force field should, in conjunction with one another, reproduce the portion of the reference CG forces that stem solely from the non-bonded AA forces. Additionally, it is interesting to note that since the AA force field used in the ZIF-8/PVDF AA-MD simulation includes electrostatic interactions (see simulation details in section 4 of the the SI), fingerprints of this interaction will also be naturally passed along to the non-bonded FM potentials within the cutoff considered (15 Å). Finally, in the context of the methodology, it is worth highlighting that the approach used to model the interactions in the ZIF-8/PVDF AA simulations is the same as the one extensively used to study several MOF/polymer interfaces.[14,15,17,18,30] As this approach has led to models in alignment with experimental results, the forces experienced by the atoms should be reasonably well depicted, which therefore enables the ZIF-8/PVDF atomistic simulations to reliably serve as a means to get data to be used as the input (CG configurations and reference forces) in the FM algorithm.

The FM algorithm was carried out using BOCS for each group of 500 configurations individually.[52] This is done because averaging the solutions coming from different groups has been observed in previous works to lead to more accurate results.[53] While defining all non-bonded potentials in the CG force field within the algorithm is important for the consistency of the overall approach set up herein (i.e., to have the conjoint CG non-bonded forces of all potentials match the AA counterpart), only the cross non-bonded potentials were further used. Before averaging the individual force profiles output by the algorithm for each MOF/polymer cross potential, outlier force values were dismissed whenever they existed. The averaged forces of each potential were then smoothened and integrated to obtain the corresponding values of potential energy.

It may be worth mentioning also that the FM algorithm can only derive data for piecewise potentials (such as the cubic splines adopted herein) in ranges of pairwise distances that are sampled within the set of CG configurations used as input. A discussion about this is made in section 5 of the SI. In the context of this work, this means that no data for the cross non-bonded potentials is output by BOCS before the minimal pair distance reported in the corresponding RDFs. Thus, values of force and energy at pair distances inferior to the one for which sampling and therefore output data exists within the FM algorithm were obtained by extrapolation. For that purpose, an exponential decay function was fit using as basis points in the onset of the interval for which data coming from FM existed for each of the non-bonded potentials. A smoothening was made in the region where originally existent data meets data coming from the extrapolation to avoid noise. The tabulated cross non-bonded potentials were then set up instead of the MARTINI counterparts, forming thus a new, hybrid MARTINI/FM CG force field for ZIF-8/PVDF.

The MD simulation of the ZIF-8/PVDF CG slab system was then repeated using the hybrid MARTINI/FM force field to assess the underlying polymer penetration that the new force field predicts. The density profile along the direction normal to the ZIF-8 slab for the last attained configuration after a total simulation time of 61M timesteps was built following the same approach as for the full MARTINI force field and is shown in fig. 2(b). It is possible to see a significant difference upon contrasting this plot with the density profile obtained when using the full MARTINI force field (fig. 2(a)). The polymer density (dark red) now exhibits more modest values in the region of overlap with ZIF-8 and eventually reaches zero, leaving a ≈32 Å thick region in the z direction of the simulation domain where only ZIF-8 bead density (light blue) is non-zero. This means that the MOF phase is empty of polymer throughout this region. From a qualitative point of view, it is expected that the result attained using the hybrid MARTINI/FM force field is better aligned with the experimental constatation of an increase in free volume upon incorporating ZIF-8 fillers into a polymeric matrix. Notably, the simulation time considered is much larger than the one used when assessing the full MARTINI force field, and it is beyond sufficient for observing fluctuations of instantaneous thermodynamic properties around a constant mean value, which hints that the system is well equilibrated. This shows that the result shown in fig. 2(b) is indeed robust. The better alignment with experimental results led to choosing the hybrid MARTINI/FM force field over the full MARTINI force field initially assembled for further investigating ZIF-8/PVDF at the nanoparticle level. The complete parameters for the MARTINI potentials composing the hybrid MARTINI/FM force field and a plot of the FM potentials can be found in section 6 of the SI. Potential tables for the non-bonded FM potentials featured can be found elsewhere.[54] It is worth noting that no pressure correction was needed within the hybrid MARTINI/FM force field for simulating the ZIF-8/PVDF slab system at ambient (P,T) conditions.

*Simulations for ZIF-8/PVDF systems at the nanoparticle level*

To study ZIF-8/PVDF at the nanoparticle level, cubic and rhombic dodecahedron morphologies were selected, as these are experimentally reported for ZIF-8.[55,56] One cubic and two rhombic dodecahedron nanoparticles with different sizes were considered herein. A configuration for each of them was built by creating a ZIF-8 supercell in the CG resolution indicated in fig. 1(a) and cutting it on the crystallographic planes underlying the given morphology. Naturally, undercoordinated beads of types 1 and 2 exist at the surface. Further visualizing the nanoparticles allowed seeing that the surface is formed by 4 and 6 membered rings (4-MR and 6-MR), featured in the crystalline structure of ZIF-8, that remained intact. Following this observation, undercoordinated ligands were added or removed at the surface so that the total amount of beads of each type ultimately follows the stoichiometry of ZIF-8 (Zn[2mIm]$_2$) and in a fashion where all the surface 4- and 6-MR windows, individually, are equivalent throughout the surface of all three nanoparticles. More specifically, Zn beads forming the 4- and 6-MR windows at the surface are terminated with undercoordinated ligands in alternating fashion so that each undercoordinated bead type 1 is surrounded by two other beads type 1 that are bonded to an undercoordinated ligand and vice-versa. A figure illustrating this termination scheme is shown in section 7 of the SI to aid the visualization. Finally, the largest distance between a ZIF-8 bead and the center of mass of the nanoparticle was calculated for each of the three nanoparticles as a means to measure the nanoparticle size. Values of 69, 62 and 28 Å were found for the cubic and two rhombic dodecahedron particles, which are respectively named herein as CB, RD and SRD. While these values imply sizes smaller than the ones most commonly studied in experimental works (nanoparticle radius ranging from 30 to 300 $\mu$m),[55,57-60] they are similar to the smallest sizes reported (5 nm radius).[59] A configuration for each of the three nanoparticles is made available elsewhere.[54]

In order to build a ZIF-8/PVDF configuration for each of the three nanoparticle systems, a configuration containing 68 unwrapped PVDF chains coarsened according to the schematics in figure 1(b) was propagated 5 times in each direction. Exceptionally, no PVDF configuration was replicated in the middle of the simulation box, slot at which the nanoparticle was placed instead. No overlap between the MOF and polymer phases was introduced when creating the configurations for the three nanoparticle systems. Rather, empty space exists between some of the PVDF chains as well as between the PVDF chains and the ZIF-8 nanoparticle. As the procedure is identical for all three nanoparticle systems, the amount of polymer beads comprised in the simulation box is the same in the three cases. Naturally, since the size of the nanoparticles are all different from one another, the compositions of the three ZIF-8/PVDF systems are also different. Ultimately the total amount of beads in the simulation box for the CB, RD and SRD ZIF-8/PVDF nanoparticle systems were 69384, 68592 and 64560, respectively.

Starting from the initial configuration built, each of the three ZIF-8/PVDF nanoparticle systems underwent a CG-MD simulation with the same setup. At a first moment, an MD-based approach similar to the one used for the CG ZIF-8/PVDF slab system was deployed for the system at the nanoparticle level aiming to reach a more suitable configuration. Details are found in the SI (section 8). Starting from such configuration, a CG-MD simulation was made using the NPT equations of motion for the bead dynamics and considering a timestep of 20 fs. Target temperature and pressure values of 300 K and 1 atm, respectively, were used. The equilibration length was not pre-established, but rather adapted according to the needs of each nanoparticle system. Notably, a very large simulation time was required for the CB and RD systems compared to the SRD one in order to start observing fluctuations of instantaneous values of potential energy and volume around an approximately constant equilibrium value (see discussion and plots in section 8 of the SI), point at which configurations for studying the systems started to be collected. Given that the equilibration in each of the three nanoparticle systems is concomitant to a progressive penetration of polymer into the nanoparticle, the fact that the SRD system requires a shorter equilibration could be linked to the fact that it features a smaller nanoparticle size to be filled by the polymer. Ultimately, configurations stemming from the range 66M-78M, 65M-81M and 19M-40M timesteps were collected for further studying the polymer structuration and penetration in the CB, RD and SRD ZIF-8/PVDF systems.

*Analysis of the ZIF-8/PVDF simulations*

For each ZIF-8/PVDF system, bead density profiles, 111 ADFs for the polymer (bead type indexing in accordance with fig. 1(b)), (ZIF-8)-(PVDF) RDFs for end-chains PVDF beads and amount of polymer beads lying within the nanoparticle domain were computed. The density profiles computed concern ZIF-8 and PVDF bead densities for the whole simulation box domain as well as PVDF beads at the surface of the ZIF-8 nanoparticle. Density profiles for

ZIF-8 and PVDF beads in the simulation domain for each ZIF-8/PVDF system were built based on the number of beads sitting at a given distance from the center of mass (COM) of the nanoparticle. All bead types for ZIF-8 and PVDF phases were considered, regardless of their type. The 111 ADFs for the polymer were evaluated locally based on the distance between the central atom and COM of the nanoparticle. Regions comprising a range of distances of size ≈5 Å were considered. The (ZIF-8)-(PVDF) RDFs were calculated globally, i.e. without distinguishing the polymer beads by how deep in the ZIF-8 nanoparticle they are. Further specific technical details about the calculation of the bead density profiles in the simulation domain, ADFs and RDFs can be found in the SI (section 8).

For building polymer density profiles at the nanoparticle surface, points of assessment were defined in the ZIF-8 surface for each of the three systems aiming to investigate the polymer structuration near the vertices, edges and faces of the nanoparticles. Each of these can be regarded as a different local environment. Upon visualizing the nanoparticle configurations before adding polymer, it is possible to see that the vertices of the CB, RD and SRD nanoparticles are all marked by either a 4- or 6-MR window. In face of this, the point of assessment for each vertex was defined at the center of mass of the undercoordinated ligands featured in the respective nearby 4- or 6-MR window. Following this definition, points of assessment for edges and faces can be defined at the bisector of the edges and center of the faces of the polyhedron formed by connecting the points of assessment of the vertices according to the nanoparticle morphology. The PVDF bead density profiles at the vertices, edges and faces were built as a function of the distance between the beads and their point of assessment ($d$) and of the smallest angle formed between the bead, the point of assessment and the center of the nanoparticle ($\alpha$). A figure to assist with the visualization of these parameters can be found in the SI (section 8). The density profiles were built considering beads lying within a distance of 12 Å, 12 Å and 5 Å from the point of assessment for the CB, RD and SRD ZIF-

8/PVDF systems respectively. These thresholds were established taking into consideration the size of the nanoparticles as well as aiming to avoid overlap between environments. Interestingly, the polyhedra underlying the two nanoparticle morphologies considered in this work have a symmetry intrinsic to their geometry which is kept herein as all 4- and 6-MR windows featured in the nanoparticle surface are terminated following the same scheme (see previous discussion). A cube features 8 vertices, 12 edges and 6 faces while a rhombic dodecahedron features 14 vertices, 24 edges and 12 faces. All vertices, edges and faces in the cubic geometry are equivalent environments. The same can be said about the edges and faces of a rhombic dodecahedron. However, in the latter case, two distinct group of vertices exist: one in which the edges forming the rhombus face meet at an angle of 70.53°, named herein group 1 (g1), and another in which the edges meet at an angle of 109.47°, group 2 (g2). A total of 6 g1 and 8 g2 vertices exist. In face of this equivalency, the density profiles are averaged for equivalent environments (vertices, edges, faces of the cube and vertices g1, g2, edges and faces of the rhombic dodecahedron) when assessing the polymer structuration at the surface level.

Figure 3 shows the three nanoparticles together with points of assessment for four vertices (dark green) and two points of assessment for an edge and face (light green). Square and rhombi faces featured in the cubic and rhombic dodecahedron nanoparticle morphologies are highlighted in dark green to aid the visualization. The spatial arrangement of ZIF-8 beads in each distinct environment found within 12 Å of the points of assessment at the vertices, edges and faces of the rhombic dodecahedron and cubic morphologies are also shown in the figure. The RD and SRD nanoparticles share the same structuration of ZIF-8 beads in the local environments, except that the actual bead density profiles for the SRD nanoparticle system were computed counting only beads lying within a 5 Å distance from the points of assessment, as previously mentioned. It is possible to identify in fig. 3 that all the environments within the two morphologies feature one or more 4- and 6-MR windows. The position of the nearest 4-

and 6-MR windows relative to the point of assessment of the different environments can be identified by the distance between the point of assessment and the window center together with the smallest angle formed between the center of the nanoparticle, the point of assessment and the center of the window. The list of these distances and angles can be found in table S3 of the SI together with further details on the calculation of polymer density profiles at the ZIF-8 surface level.

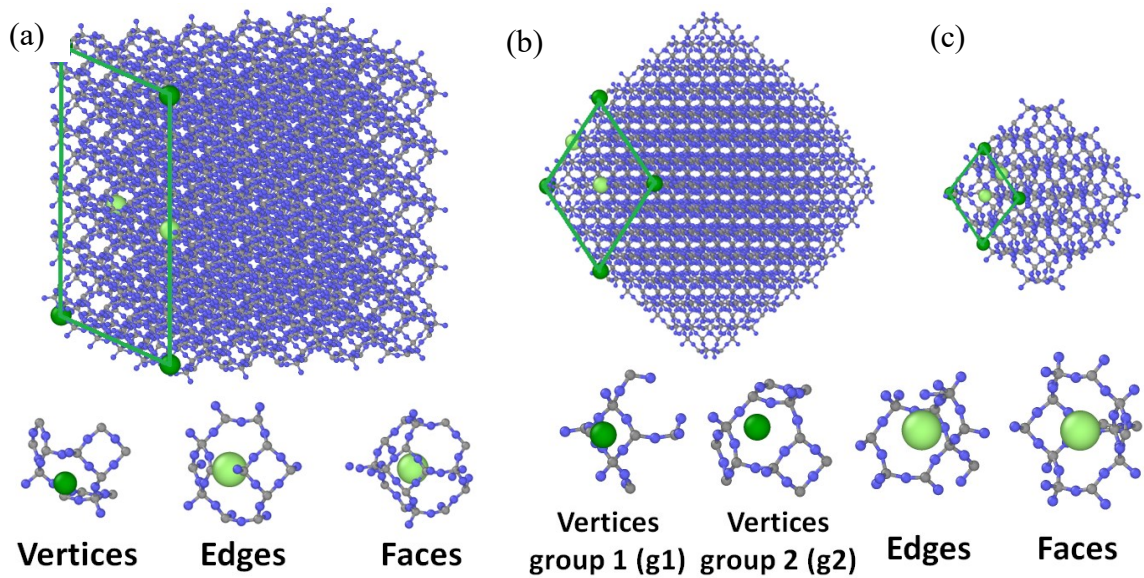

Figure 3. Illustration of the (a) CB, (b) RD and (c) SRD ZIF-8 nanoparticles considered in the investigation, each containing four points of assessment of vertices (dark green) and two (light green) for the edges and faces each. Rhombi and cubic faces are highlighted in dark green to aid the visualization. The gray and blue beads are beads type 1 and 2, respectively. The ZIF-8 beads lying within distance ≤ 12 Å from the points of assessment of vertices, edges and faces are also shown. Within this distance, it is possible to see most of the 4- and 6-MR windows featured in each environment, with some windows being cut in some cases due to the cutoff.

Finally, the amount of polymer beads lying in the nanoparticle domain for the CB ZIF-8/PVDF system was computed using an in-house python code based on the average amount (within the set of configurations considered) of polymer beads lying within a cubic region of length 97.6

Å centered in the center of mass of the CB nanoparticle. For the RD and SRD ZIF-8/PVDF systems, the amount of polymer beads in the nanoparticle domain was estimated by averaging the average number of beads lying within the circumscribed (radius ≈62 Å and ≈48 Å) and inscribed spheres (radius ≈28 Å and ≈24 Å) of the RD and SRD nanoparticles, respectively. Notably, since all these nanoparticles have different sizes, it is necessary to normalize the number of PVDF beads in the ZIF-8 domain by the nanoparticle size for a fair comparison that allows to evaluate influence of nanoparticle morphology and shape since the influence of composition is ruled out. The number of ZIF-8 beads composing the CB, RD and SRD were respectively considered as normalizing factors.

For the sake of having further reference for analyzing the results, the AA-MD simulation of the ZIF-8/PVDF slab system used to parametrize the FM potentials of the hybrid MARTINI/FM force field was also employed to compute (ZIF-8)-(PVDF) RDFs at the CG level and a density profile along the direction normal to the ZIF-8 slab. The density profile was calculated without coarsening the AA configurations while, naturally, these were coarsened for computing the RDFs for a more straightforward comparison with the results in the CG level. In this context, the atomic density for the bulk AA PVDF was computed to confirm if the density of the polymer outside the MOF/polymer overlap region in the AA ZIF-8/PVDF slab system converges to it. Further details of these calculations can be found in section 9 of the SI. The 111 ADF obtained for the bulk CG polymer (using the optimal force field attained in the optimization of the PVDF bond potentials) is also used as reference to compare with the local 111 ADFs built for ZIF-8/PVDF at the nanoparticle level.

*Studying the ZIF-8/PVDF systems at the nanoparticle level under $CO_2$ loading*

Finally, the three ZIF-8/PVDF nanoparticle systems were investigated under $CO_2$ loading to evaluate if the model indeed predicts that the ZIF-8/PVDF composite has a superior volume for gas adsorption compared to the PVDF matrix, as implied by the experimental results.[45-48] To do this investigation, the gas was considered in CG resolution with a single bead representing each molecule, and a force field for ZIF-8/PVDF/$CO_2$ was assembled by introducing (ZIF-8)-($CO_2$), (PVDF)-($CO_2$) and ($CO_2$)-($CO_2$) MARTINI potentials to model the guest-host and host-host interactions. These potentials were fully determined within MARTINI 3 by classifying the $CO_2$ beads as TN3 and using the bead flavor classifications for ZIF-8 and PVDF discussed above.

To study the gas adsorption, one configuration for each binary ZIF-8/PVDF nanoparticle system was randomly selected from the respective production run and used as initial configuration for a hybrid Monte Carlo (MC)/MD simulation. The MC moves consisted solely of $CO_2$ insertions and deletions. The simulation details can be found in the SI (section 10). Once the equilibrium adsorbed amount of molecules is reached, an MD simulation is performed using the NPT equations of motion for the dynamics of all beads. Target pressure and temperature were set to 1 atm and 300 K, respectively. The timestep was 10 fs. A set of 1200 configurations for each equilibrated ZIF-8/PVDF nanoparticle system were collected and used to build density profiles throughout the entire simulation domain for ZIF-8, PVDF and $CO_2$ beads. The latter calculation was made following the same guidelines as for the binary system, as mentioned previously and explained in more detail in section 8 of the SI, with the addition that the density profile for the gas was also computed in this case. Additionally, the number of gas molecules lying within the ZIF-8 nanoparticle domain for all systems was also computed for the ZIF-8/PVDF/$CO_2$ systems following the same methodology as for computing the number of PVDF beads inside the nanoparticle domain for the binary systems. In this matter, it may be worth stressing that since the composition of the CB, RD and SRD ZIF-8/PVDF systems are all different, normalization of the values by the number of ZIF-8 beads was also made here aiming to isolate the effect of nanoparticle morphology and size.

## Results & Discussion

To begin the analysis, the ZIF-8 (light blue) and PVDF (dark red) bead density profiles in the overall simulation domain for each of the three ZIF-8/PVDF nanoparticle systems as a function of its distance from the center of the nanoparticle are shown in figures 4(a)-(c). The vertical gray dashed lines in figures 4(a)-(c) have equation x = $a$, where $a$ is the bin sitting at the largest distance from the center of the nanoparticle and featuring a non-zero ZIF-8 bead density value. In other words, $a$ serves to mark the end of the MOF nanoparticle phase.

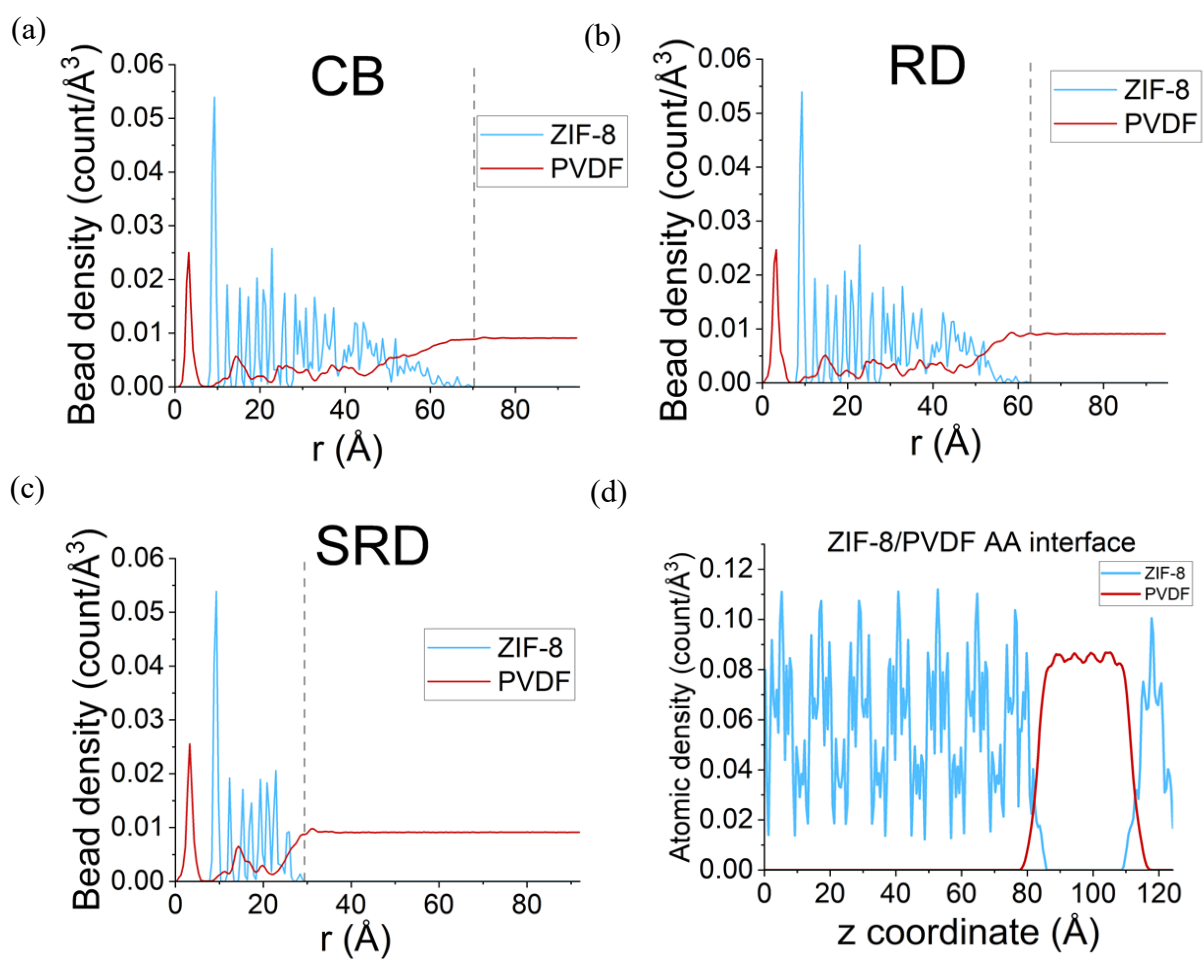

Figure 4. ZIF-8 (light blue) and PVDF (dark red) bead density values for the (a) CB, (b) RD and (c) SRD ZIF-8/PVDF nanoparticle systems, studied at the CG level, and (d) an AA ZIF-8/PVDF system. The density profiles at the nanoparticle level were calculated considering the distance between the bead and the center of the nanoparticle while the profile for the AA ZIF-8/PVDF slab system was computed along the direction normal to the ZIF-8 slab.

As it can be seen by comparing figures 4(a)-(c), the density profiles obtained for the CB, RD and SRD ZIF-8/PVDF systems in the bulk-like region of the ZIF-8 nanoparticle have very similar form, suggesting that the influence of nanoparticle shape on the polymer structuration in the overall simulation domain is not straightforwardly evident. Complementary to the information portrayed in the PVDF bead density profiles, the amount of polymer beads lying within the nanoparticle domain for the CB, RD and SRD ZIF-8/PVDF systems may thus offer valuable insights. These values are $0.743 \pm 0.004$, $0.713 \pm 0.002$ and $0.427 \pm 0.001$ PVDF beads per ZIF-8 bead, respectively. The uncertainties in the values correspond to the standard error estimated at 95% confidence. This points towards a trend of CB > RD > SRD for polymer adsorption in the overall nanoparticle domain. Interestingly, when using the bead density profiles to compute the amount of polymer lying within $r \leq 20.3$ Å, average values of 91, 87 and 109 polymer beads are found for the CB, RD, SRD ZIF-8/PVDF systems, respectively, which suggests that the polymer reaches more easily and/or more intensively the innermost region of the SRD than the CB than the RD nanoparticle (SRD > CB > RD trend). Putting this together with the lower amount of PVDF beads in the whole nanoparticle domain in the SRD ZIF-8/PVDF system compared to the RD one, the explanation, by exclusion, is that the polymer has more difficulty adsorbing near and/or at the nanoparticle surface in the case of smaller sized nanoparticle. Thus, when adsorption in the overall nanoparticle domain is more important for a given application, it could be concluded that smaller sized nanoparticles are more advantageous by reasoning that a more intense polymer penetration would imply less free space for the gas. On the other hand, the values obtained in the CB and RD ZIF-8/PVDF systems are relatively close, and thus a more careful investigation would be required to evaluate whether this is relevant from the application point of view.

The atomic density profile built for the AA ZIF-8/PVDF slab system along the direction normal to ZIF-8 slab surface is shown in figure 4(d) for further reference. It may be worth highlighting that the very large difference in magnitude in density values (y-axis) observed between the atomic (figure 4(d)) and bead (figures 4(a)-(c)) density profiles should indeed exist, as diminishing the amount of minimal structural units composing the system is the very purpose of coarse graining and therefore the values *per se* should not be directly compared. However, the profiles can be qualitatively compared. It is possible to see that the polymer penetration in the CG systems at the nanoparticle level is significantly deeper than what is observed in the AA ZIF-8/PVDF slab system, as the red curve extends all throughout the MOF phase. Yet, despite this difference, the plots shown in figure 4(a)-(c) and those shown in figure 4(d) do share some features. Firstly, upon respectively comparing the atomic and bead density values obtained for the bulk AA and CG PVDF (0.085 and 0.009 count/Å$^3$) with the values attained outside the overlap region, it is possible to conclude that convergence to PVDF bulk density values outside the overlap region happens right away in all cases, suggesting that there is no long-range influence of ZIF-8 on polymer density. Secondly, it is possible to see that, except for the unphysical steep peak near the nanoparticle center, PVDF density in the region of overlap with ZIF-8 is always less than the bulk value in all cases. The very large peaks at low distances in figures 4(a)-(c) are an artifact of the methodology used to compute the density profiles, which involves dividing the total amount of beads by the volume of the region in which the bead count is made: as the volume of the regions used at low r values is very small, the result of the division unphysically increases (see section 8 of the SI for details on the methodology).

Reconsidering the bead density profile for the CG ZIF-8/PVDF slab system, built for preliminary evaluating the CG force field and shown in fig. 2(b), allows for further valuable insights. It is possible to see that the CG ZIF-8/PVDF slab system resembles more the AA ZIF-

8/PVDF slab system than the CG ZIF-8/PVDF systems at the nanoparticle level. Despite the slightly larger polymer penetration in the CG compared to the AA ZIF-8/PVDF slab system, it can be seen that the polymer penetration at the former is much more limited compared to the one observed for all three systems in the nanoparticle level even though the total simulation timescales are comparable. Given that the force field modeling the interactions is the same in the CG ZIF-8/PVDF slab and nanoparticle systems, these results suggest that depicting the proper polymer penetration may not be merely a matter of simulation resolution (CG versus AA), but also of simulation scale (nanoparticle versus surface slab). Vertices and edges of the nanoparticle seem to facilitate the penetration of PVDF chains compared to the faces, which would resemble the surface slab case better. This conclusion is particularly corroborated by the fact that the crystallographic plane in which ZIF-8 was terminated in the AA and CG ZIF-8/PVDF slab systems considered herein ([011] plane) is equivalent to the planes forming the faces in the RD and SRD nanoparticles specifically in terms of environments. This means that, in this case, changes in penetration propensity cannot be justified by a difference in the ZIF-8 structure.

Notably, despite the importance of vertices and/or edges for properly depicting the penetration within the ZIF-8/PVDF systems investigated herein, it is not possible to draw conclusions on the mechanism through which the PVDF penetration in ZIF-8 occurs from these simulations. Regardless, hypotheses can be made. Although penetration can occur through vertices, edges or faces, it could be that the probability of polymer beads passing through the vertices or edges is larger. In this matter, given that all the environments (vertices, edges and faces) feature both 4- and 6-MR windows, the larger PVDF chain penetration at vertices and/or edges is likely tied to the angle associated to these environments within the nanoparticle morphology as well as, possibly, to the specific ZIF-8 bead arrangements (i.e., amount of 4-MR and 6-MR windows and how they are placed in space). Alternatively, or in conjunction with this hypothesis, polymer penetration may be affected by a collaborative effect of PVDF chains already inside the ZIF-8 phase, which may facilitate the penetration of other chains lying near the surface via a polymer/polymer attractive interaction. Notably, in this case, the main factor allowing to fairly depict the polymer penetration would not be the presence of vertices and edges *per se*, but rather the possibility of having shorter inter-chains distances within the ZIF-8 phase.

The PVDF bead density profiles built near vertices, edges and faces of the nanoparticle surface are presented in Figure 5 and in the SI. Results are in alignment with polymer penetration occuring in all the environments for the three nanoparticle systems, which agrees with the previous results and discussion. More specifically, all of them evidence non-zero bead density values near the 6-MR windows featured in each environment (marked by a given (d, $\alpha$) value, as indicated in table S3) as well as at (d, $\alpha$) lying inside and outside the ZIF-8 nanoparticle domain. This means that there is a continuous region populated by the polymer that comes from the polymer phase outside the overlap region into the nanoparticle passing through the 6-MR windows. This could be associated with a path for the polymer to penetrate into the ZIF-8 nanoparticle.

As an example case, the PVDF bead density profile associated with the point of assessment of the vertices of the CB ZIF-8/PVDF system is shown in figure 5. In the figure, the x and y axis show the values of $\alpha$ and d associated to each point in space, while the color scale indicates the magnitude of the linear density (in bead/Å units) associated with each given combination (d,$\alpha$) (see the SI for more details on the calculation of polymer bead densities at the nanoparticle surface level). The center of the 6-MR window featured in the vertices of the CB ZIF-8 nanoparticle lies at d = 2.49 Å and $\alpha$ = 6.46°, as indicated in fig. 5. By superimposing the position of the point of assessment relative to the 6-MR window and to the ZIF-8 nanoparticle shown in fig. 3, it is possible to identify a set of points (d,$\alpha$) that would correspond to a path for polymer beads going from the outside to the inside of the ZIF-8 nanoparticle. The beads that are outside the nanoparticle and beyond the point of assessment, should have large values of $\alpha$ and decreasing values of d as the bead approaches the point of assessment. As the polymer bead reaches a position in between the point of assessment and the nanoparticle, the values of $\alpha$ shift from large to small ones. From there, the crossing through the 6-MR window implies an increase in the distance from the point of assessment. For example, in figure 5, the PVDF beads outside the nanoparticle domain yet near the point of assessment would be at d∈[0,2]Å and $\alpha$∈[120,180]°. As they cross the point of assessment and find themselves in between it and the nanoparticle center, the values of $\alpha$ would now lie within [0,30]° and similarly low distances (d∈[0,2]Å). Non-zero PVDF density values at these regions of the plot suggest, then, a possible polymer penetration path. Analysis of the other density profiles presented in the SI leads to similar conclusions as those made for fig. 5. On the other hand, the 4-MR windows do not seem to be associated with a path for polymer penetration, which is briefly discussed in the SI. Indeed this result is aligned with the fact that the diameter of the 6-MR window is larger than that of the 4-MR, which should thus facilitate the polymer penetration.

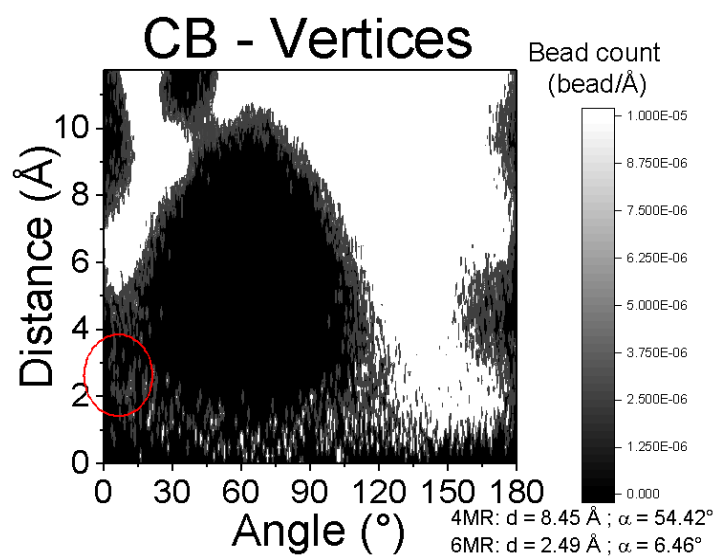

Figure 5. PVDF bead density profile built around the point of assessment of the vertices of the CB ZIF-8/PVDF nanoparticle system. The angle ($\alpha$) formed between the center of the nanoparticle, the point of assessment and the PVDF bead is shown in the x axis, the distance (d) between the bead and the point of assessment is shown in the y axis, and the color scale depicts the value of the polymer density at a given (d,$\alpha$). The value of (d,$\alpha$) associated to the center of the 4- and 6-MR windows featured in this environment are shown below the plot. A red circle indicates the region near the center of the 6-MR window.

The 111 ADFs for the bulk CG polymer and for the polymer in the ZIF-8/PVDF nanoparticle systems in specific regions of the simulation domain are shown in figure 6. Bead types are indexed following the definition in figure 1(b). In the ZIF-8/PVDF figures, the ADFs concern different ranges of distances (in angstroms) between the central bead forming the 111 angle and the center of the nanoparticle. Comparing ADF profiles built for different ranges allows evaluating changes in polymer conformation as a result of polymer beads being inside or outside the nanoparticle domain. On this matter, it is worth noting that the ZIF-8 beads sitting the farthest from the center of the nanoparticle in the CB, RD and SRD ZIF-8/PVDF systems are on average at distances of ≈70 Å, ≈63 Å and ≈29 Å, respectively. Upon comparing figure 6(a) to figures6(b)-(d), it is clear that the 111 ADF profile for PVDF changes from the one adopted by the bulk polymer as soon as the central atom reaches the nanoparticle domain in all three ZIF-8/PVDF systems. The change in chain structuration should be possible due to the relative flexibility of the polymer and should reflect both the pore topology it adapts to as well as the positioning of the ZIF-8 beads the polymer most favorably interacts with.[18] On the other hand, the 111 ADF profiles outside the overlap region match the ones predicted for the bulk CG polymer immediately after crossing the nanoparticle surface in the CB, RD and SRD ZIF-8/PVDF systems. This reinforces the previous conclusion made in the context of the density profile about there being no long-range effect of the ZIF-8 nanoparticle on PVDF structure.

Ultimately, the similarity of the ADFs inside the nanoparticle region in the CB, RD and SRD ZIF-8/PVDF systems suggest the change in conformation is the same in the three cases, implying that morphology and size play a negligible role in the resulting polymer conformation within the nanoparticle domain. The subtle differences observed in the form of the 111 ADFs inside the nanoparticle domain (e.g. black curve) in the case of SRD ZIF-8/PVDF system compared to the other two systems are believed to be of statistical nature.

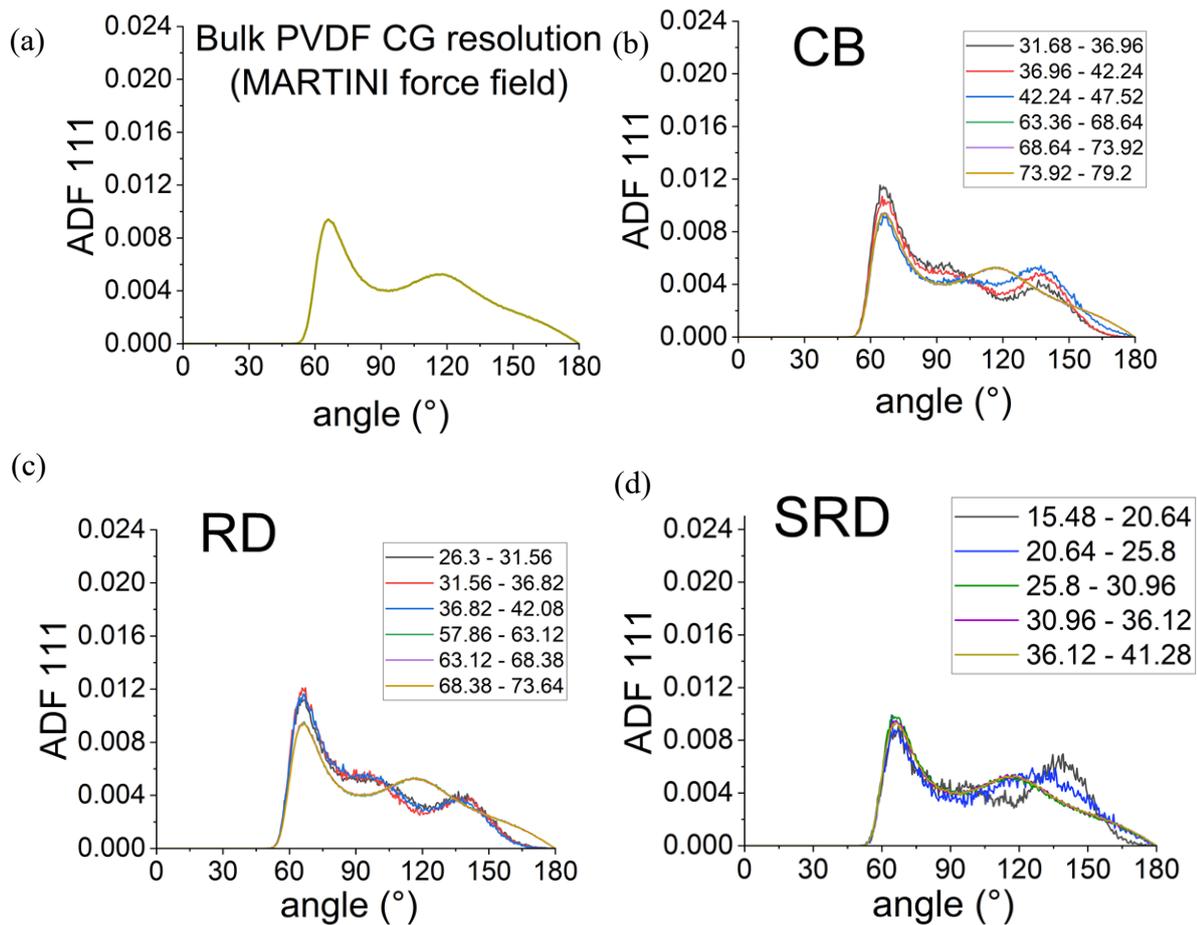

Figure 6. 111 ADF profile for the (a) bulk CG polymer, (b) CB, (c) RD and (d) SRD ZIF-8/PVDF systems. The three latter plots distinguish 111 ADF profiles based on the underlying distance between the central bead of the angle and the center of the nanoparticle, aiming to elucidate if the chain structure changes with the depth of polymer penetration. The range of distances used to compute each curve is indicated in the caption. Only a few selected relevant curves are shown.

The (ZIF-8)-(PVDF) RDFs built for pairs involving polymer beads at the chain end for the coarsened AA ZIF-8/PVDF slab system and for the CB, RD and SRD ZIF-8/PVDF systems are shown in figure 7(a) and 7(b)-(d), respectively. The pairs of bead types corresponding to each curve are shown in the caption of the figures following the indexes defined in figures 1(a) and (b). While the form of the RDF profiles for the coarsened AA ZIF-8/PVDF slab system may be slightly different than their counterparts in the nanoparticle systems, it is possible to see that they strongly resemble one another, practically matching the pair distances in which peaks are observed. The main difference is in the magnitude of values of the functions, which should be tied to the significantly larger polymer penetration observed at the nanoparticle level, implying a larger number of neighbors compared to the coarsened AA ZIF-8/PVDF slab system. The similarity of the results shown in figure 7(a) and in 7(b)-(d) advocates in favor of the ability of the hybrid MARTINI/FM force field developed herein in reproducing the structure prescribed in the AA ZIF-8/PVDF slab model for the binary system. Such ability should stem from the fact that the reference forces associated to the cross interactions are well reproduced by the FM cross potentials within the FM algorithm setup. Finally, comparing figures 7(b)-(d) allows concluding that there is negligible influence of nanoparticle size and shape in the polymer structure.

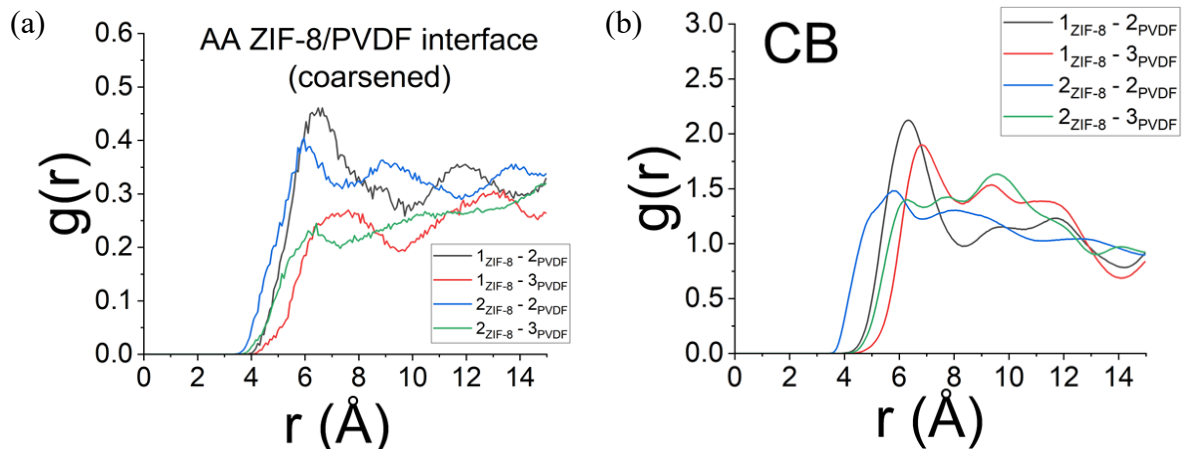

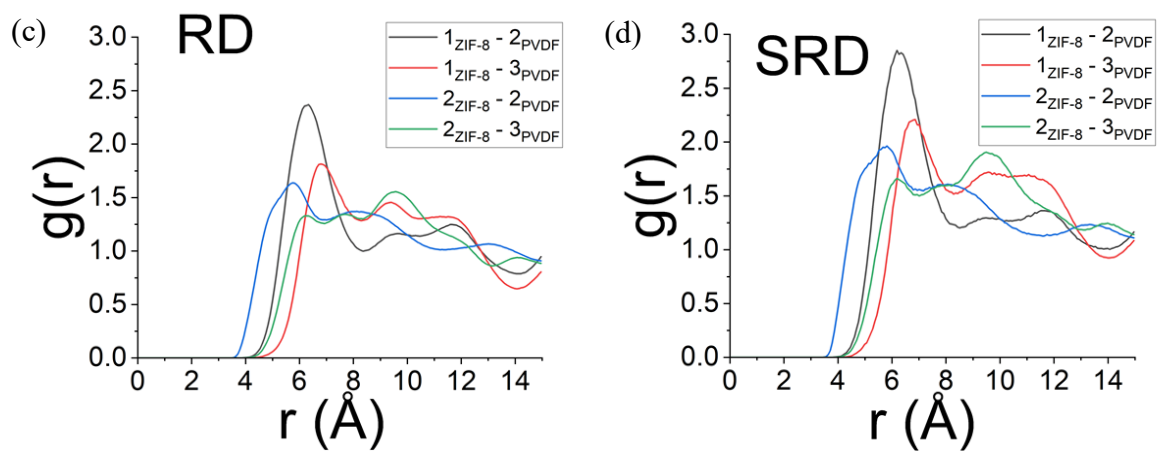

Figure 7. RDFs for (ZIF-8)-(PVDF) beads calculated for the (a) coarsened AA ZIF-8/PVDF slab system, (b) CB, (c) RD and (d) SRD ZIF-8/PVDF systems. Bead types of ZIF-8 and PVDF follow the definition made in fig. 1(a) and (b), respectively. Only RDFs featuring bead types at PVDF chain ends are shown.

One could argue that the fact that the polymer penetrates the entire length of the ZIF-8 phase in the nanoparticle systems may cast a shadow of doubt on whether the force field will indeed be able to depict the gas adsorption experimentally observed in ZIF-8/PVDF membranes.[45-48] The study of the $CO_2$ loaded ZIF-8/PVDF systems at the nanoparticle level can shed light on the topic. The equilibrium amount of $CO_2$ molecules adsorbed in the simulation domain at ambient (T,P) conditions are $445 \pm 37$, $415 \pm 8$ and $189 \pm 11$, respectively for the CB, RD and SRD ZIF-8/PVDF systems, where the uncertainty corresponds to the standard error at a 95% confidence. Figure 8(a) and (b) show the $CO_2$ bead density profiles for the $CO_2$ loaded CB ZIF-8/PVDF system obtained at equilibrium at ambient (T,P) conditions. Figure S12 in the SI shows analogous plots for the RD and SRD ZIF-8/PVDF systems. The figure in the right hand side corresponds to a zoom-in version of the one in the left hand side aiming to highlight the values of $CO_2$ density attained in the PVDF phase outside the overlap region. The relatively large $CO_2$ density values in the ZIF-8 phase shown in fig. 8(a) prove that polymer penetration is not incompatible with gas adsorption. More importantly, the higher $CO_2$ densities in the ZIF-

8 nanoparticle domain compared to the PVDF phase outside the MOF/polymer overlap region lead to conclude that the CG model is aligned with the increase in gas adsorption experimentally observed upon incorporating ZIF-8 fillers into a PVDF matrix.

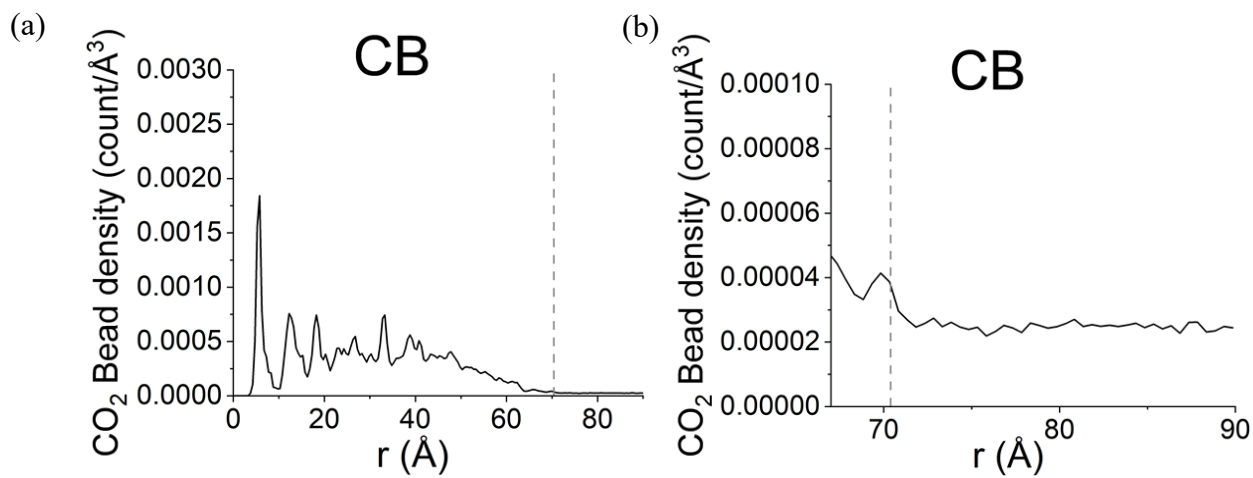

Figure 8. $CO_2$ bead density profile (a) along the simulation domain and (b) at and near the PVDF phase outside the MOF/polymer overlap region for the CB ZIF-8/PVDF system.

The calculated amount of $CO_2$ in the ZIF-8 nanoparticle domain at equilibrium may offer further insights on the influence of nanoparticle size and shape on gas adsorption. These values are $32.2 \pm 0.8$, $29.4 \pm 0.2$ and $33.5 \pm 0.8$ mg $CO_2$/g ZIF-8 for the CB, RD and SRD ZIF-8/PVDF systems. The uncertainty corresponds to the standard error at a 95% confidence in all cases. The differences in the values suggest that the cubic morphology is more favorable to $CO_2$ adsorption compared to the rhombic dodecahedron, while smaller sizes also seem to contribute to a larger amount of molecules adsorbed within the nanoparticle size range considered herein. More specifically, the equilibrium amount of gas adsorbed follows the order SRD > CB > RD. The higher capacity for adsorbing gas displayed by the SRD ZIF-8/PVDF nanoparticle system is expected from the straightforward reasoning made above: since the amount of polymer adsorbed into the nanoparticle doming is relatively smaller for the SRD, more free space for gas adsorption should remain. However, the result observed for the CB versus the RD ZIF-8/PVDF system cannot be explained by this reasoning. Furthermore, the difference in polymer penetration observed between these two systems (0.743 ± 0.004 versus 0.713 ± 0.002 polymer beads per ZIF-8 bead) is of much lower magnitude compared to the one observed when comparing the RD and SRD ZIF-8/PVDF system (0.713 ± 0.002 versus 0.427 ± 0.001 polymer beads per ZIF-8 bead), which suggests this is not the source of the difference in gas adsorption in the former case. As in the present work, the more advantageous gas adsorption observed in the CB nanoparticle geometry than in the RD may be regarded as a finding on its own. It is worth mentioning in this context that, since the bulk $CO_2$ adsorption should not be affected by the morphology of the nanoparticle, the difference should come from gas adsorption at the surface level. Since the cubic nanoparticle has a larger surface-to-volume ratio (≈7.0% for the CB vs ≈6.2% for the RD ZIF-8/PVDF systems, respectively), it may be that more surface sites are available for $CO_2$ adsorption in the former for this reason.

## Conclusion

A MOF/polymer composite was modeled at the mesoscopic scale for the first time aiming to unveil the influence of nanoparticle morphology and size in the polymer structuration and in $CO_2$ adsorption. To this end, a hybrid coarse grained force field was developed by combining MARTINI 3 potentials to model intraphase MOF/polymer interactions with FM potentials to model interphase interactions. The idea underlying this hybrid force field strategy is that the most relevant interactions in governing the composite interface structure (i.e., those between the MOF and the polymer) are treated with higher accuracy. This is achieved by relying on a chemically specific approach that allows to converge to potentials that most optimally reproduce the reference forces stemming from these interactions at the all-atom level. This approach allows to overcome the excessive MOF/polymer attraction predicted by a full MARTINI 3 force field that was tested at a preliminary stage.

Three ZIF-8 nanoparticles were considered in the investigation: a cubic nanoparticle of radius ≈69 Å and rhombic dodecahedron nanoparticles of radius ≈62 Å and ≈28 Å. Each of these nanoparticles was embedded into a PVDF matrix to create three different ZIF-8/PVDF systems. The results were compared with the ones obtained for a smaller scale CG ZIF-8/PVDF system, featuring an infinite ZIF-8 2D slab. Polymer penetration into the MOF phase was found to be significantly more extensive in the 3D nanoparticle case than in the 2D system, despite having deployed the same force field and timescale in both simulations. Indeed, polymer chains penetrate the whole nanoparticle domain in all cases, while they only penetrate the first two pore layers in the 2D slab system. These results suggest that the vertices and/or edges featured in the nanoparticles play an important role in facilitating the polymer penetration compared to the 2D slab system investigated herein. Thus, depicting the right extent of MOF/polymer interactions may not simply be a matter of simulation resolution, but also of its scale. Alternatively, or in conjunction with it, the higher extent of polymer penetration observed at the larger scale could also be the result of a collaborative effect where polymer chains inside the nanoparticle domain attract those outside of it, facilitating their entering to the MOF pores. Investigation of polymer structuration at and near the surface of the ZIF-8/PVDF systems at the nanoparticle level suggested that the PVDF chains enter the nanoparticle domain through surface 6-MR windows. Ultimately, the smallest nanoparticle investigated has shown less polymer penetration, as less PVDF beads per ZIF-8 bead are counted in the nanoparticle domain. Differences found in polymer penetration as a function of morphology were less significant. Despite the quantitative differences in extension of polymer penetration, the PVDF bead density profiles were found to be qualitatively similar in all cases.

Analysis of angle distribution functions for PVDF allowed concluding that polymer conformation outside and inside the nanoparticle domain differ, ultimately leading to the conclusion that ZIF-8 displays a short range influence on the structure of polymer chains. No effect of nanoparticle shape and size was observed in this analysis. The ZIF-8-PVDF radial distribution functions were also similar in all cases. Finally, when tested for $CO_2$ adsorption, $CO_2$ was preferentially adsorbed in the ZIF-8 nanoparticle domain in all systems, in alignment with experimental results that show that free volume increases upon incorporating ZIF-8 fillers into PVDF. The smallest nanoparticle offered the highest gas adsorption capacity among those investigated at ambient conditions, in alignment with the lower polymer penetration extent observed in the overall nanoparticle domain. With respect to morphology, the cubic one favors $CO_2$ adsorption, possibly due to a larger surface-to-volume ratio.

The hybrid CG MOF/polymer force field development method presented herein can be applied to model other MOF-polymer pairs at the nanoparticle level. This opens the door to studying crucial aspects that govern the applicability of MOF/polymer composites to help solving environmental and industrial problems associated with gas adsorption processes.

## Acknowledgements

The authors thank the École Doctoral Sciences Chimiques Balard for funding this work. R. S. thanks the European Research Council for an ERC StG (MAGNIFY project, number 101042514). This work was granted access to the HPC resources of CINES under the allocations AD010911989R1 and A0150911989 made by GENCI.

# Supporting Information (SI)

# Coarse grained modeling of a metal-organic framework/polymer composite and its gas adsorption at the nanoparticle level

*Cecilia M. S. Alvares[a], Rocio Semino[b*]*

[a] ICGM, Univ. Montpellier, CNRS, ENSCM, Montpellier, France

[b] Sorbonne Université, CNRS, Physico-chimie des Electrolytes et Nanosystèmes Interfaciaux, PHENIX, F-75005 Paris, France

As mentioned in the main text, the Nosé Hoover equations of motion as originally developed either to sample the NVT or NPT ensemble were used for carrying out molecular dynamics (MD) simulations.[1,2] These equations will be correspondingly referred to along the text as "NVT equations of motion" or "NPT equations of motion". Unless specified differently, the damping constants for the thermostat and barostat (whenever applicable) are of 100x and 1000x the value of the timestep used for the numerical integration, respectively, in accordance with the recommendation in the LAMMPS manual.[3]

1) Atomistic simulations for bulk PVDF

In the simulations for bulk PVDF in all-atom (AA) resolution, the interactions were modeled using the DREIDING force field,[4] with the bonded potentials limited to bonds and angles (i.e., no dihedral potentials were used). The timestep for the MD was fixed to 1 fs. An initial configuration was borrowed from previous work.[5] A protocol previously designed to equilibrate the system, as it allows reaching a configuration better suitable for starting an AA-MD simulation for amorphous polymer systems at the AA level, was deployed.[6] The protocol consists of a series of MD steps performed in different temperature and pressure conditions. As protocol parameters, $P_{max}$= 50 kbar and a high temperature of 600 K was considered. Subsequent to the protocol, the dynamics was performed using the NPT equations of motion with a target temperature and pressure of 300 K and 1 atm, respectively. Configurations were then collected after fluctuations of instantaneous values of potential energy and volume suggested the system was equilibrated.

Furthermore, 10 additional initial configurations were created by running 10 independent AA-MD simulations, each contemplating 10 cycles of heating and cooling, starting from the configuration attained by the end of the previously mentioned protocol. The 10 cycles of heating and cooling used in each independent AA-MD simulation contemplate different maximum temperatures and durations. The underlying idea is that the high kinetic energy at the high temperatures featured in the cycles would allow reaching different regions of the conformational space, so that if AA-MD simulations were made using these microstates as initial ones, an ergodic sampling would be attained within a reasonable timescale by putting them all together. This is similar to what was done in previous works.[7] The NVT equations of motion were used for the dynamics during the heating and cooling cycles. Using each of the microstates attained by the end of these cycles, an additional independent AA-MD simulation was carried out using the NPT equations of motion with a target temperature and pressure of 300 K and 1 atm. Configurations were collected post equilibration in each of these 10 additional AA MD simulations.

All the configurations collected from the 11 independent AA-MD simulations were used as input for an in-house python code that coarsens the configurations according to the resolution shown in figure 1(b) of the manuscript and computes reference bond and 3-body angle distribution functions (BDFs and ADFs). Bond lengths and 3-body angle values were computed for each sequence of two and three beads that replace atoms that were bonded in the all atom resolution, respectively. From these values, histograms were constructed for each distinct sequence of two and three bead types and normalized by the total count, giving thus rise to BDFs and ADFs. Ultimately, five distinct BDFs and ADFs exist and they are shown in figure S1(a) and (b), respectively. The specific bead types in the figure are indicated in the caption and follow the definition given in figure 1(b) of the main text. Bonds between bead types 2 and 4 do not exist since the polydispersity (see figure 1(c) of the main text) of the polymer does not allow these to occur. As expected from a flexible polymeric system in which the chains can adopt a variety of different conformations, the profiles of the BDFs and ADFs in the coarse grained (CG) resolution are asymmetric and often multi-peaked.

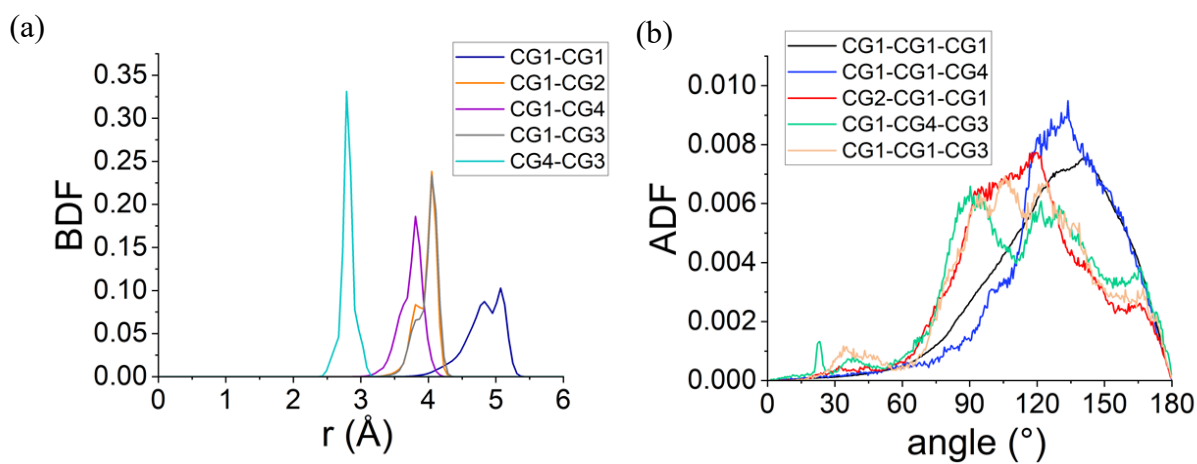

Figure S1. Normalized (a) bond and (b) angle distribution functions for PVDF in the CG resolution shown in Figure 1(b) of the main text. The figures show functions for all bonds and angles formed by a specific sequence of bead types.

2) CG-MD simulations for parametrizing the PVDF bond potentials

The CG-MD simulation for bulk PVDF starts from an initial configuration obtained from coarsening an AA configuration, unwrapping the chains and propagating it in all directions to ultimately attain a reasonable amount of structural units in the simulation box. The CG force field corresponds to the MARTINI non-bonded potentials that follow as a consequence of the bead flavor classification and five bond potentials, one for each sequence of bead types shown in figure S1(a). The $k_b$ values were initially arbitrarily set and $r_o$ were estimated from averaging the occurring values of bond lengths for the respective pair of bead types.

Aiming to get rid of any empty space between chains created during the set up of the initial configuration, a steady deformation of the simulation box up to the value of the experimentally reported density for PVDF[8] was made while the dynamics of the beads was carried out using the NVT equations of motion with a 10 fs timestep. A high target temperature (1000 K) was used aiming at a kinetic energy high enough to allow ultimately converging to conformations indeed prescribed as occurring ones in the eyes of the CG force field used. The final dimensions of the simulation box are set so that the simulation domain is cubic since no anisotropy is expected for the polymer. Afterwards, the dynamics was carried out using the NPT equations of motion considering 1 atm and 300 K as target pressure and temperature, respectively, for a total of 1 M timesteps. Subsequently, the timestep was raised to 20 fs, and, once fluctuations of instantaneous values of potential energy and volume suggested the system was equilibrated, configurations were collected and used to build the BDFs, computed and output by LAMMPS.[9,10]

The values of force constants of the bond potentials were iteratively increased or diminished up to convergence to a force field capable of reproducing the width and height of the reference

BDFs (determined in section 1). At each iteration, a CG-MD simulation relying on an updated version of the CG force field was started from the microstate reached by the end of the CG-MD simulation made using the initial force field. As this configuration does not have the artificial void created when creating the initial configuration, the protocol involving the deformation of the simulation box previously mentioned does not need to be carried out, allowing to save simulation time. Naturally, the configurations used to build the BDF using LAMMPS were collected post-equilibration during a CG-MD simulation carried out using the same equations of motion and target pressure and temperature as the initial one at each iteration. Once the BDF were accurately reproduced, the iterations were finalized at the final optimal force field was kept.

Finally, as discussed in the main text, the bulk CG PVDF density at (300 K, 1 atm) obtained using the MARTINI force field contemplating the optimized bond potentials was computed for further evaluating if the local values of PVDF density outside the region of overlap with the MOF in the ZIF-8/PVDF system match the bulk polymer value. This calculation was made considering configurations attained post equilibration in the CG-MD simulation made using the NPT equations of motion, yielding an average value of 0.009 beads/$\text{Å}^3$.

3) Details of the CG-MD simulation of the ZIF-8/PVDF slab system

An approach similar to the one deployed for the CG-MD simulation for bulk PVDF was used to simulate the CG ZIF-8/PVDF slab system. Using the initial configuration built as described in the main text, the simulation box was deformed at a steady rate in the direction normal to the ZIF-8 slab's surface (see *fix_deform* command in LAMMPS)[11] up to the point where the volume occupied by the PVDF phase would imply a density equal to the experimental one[8] for the bulk polymer assuming no penetration occurs. During the deformation, the dynamics of the beads were carried out using the NVT equations of motion at a target temperature of 300 K. A timestep of 10 fs was considered for the dynamics. Once the final volume was reached, the system underwent equilibration using the NPT equations of motion with target pressure and temperature of 1 atm and 300 K. The three dimensions of the simulation domain were allowed to change during the dynamics. The timestep was then raised to 20 fs for the equilibration and production. During the production, configurations were collected to build single-configuration density profiles for the ZIF-8 and PVDF phases along the direction normal to the surface of the ZIF-8 slab using an in-house python code, aiming to evaluate the corresponding polymer penetration. The code discretizes the z direction according to a bin size of ≈0.5 Å, and later divides the total count at each bin by the volume of the region to ultimately yield a volumetric density. No distinction on ZIF-8 and PVDF bead types was made when computing the bead density profile for the ZIF-8 and PVDF phases, respectively.

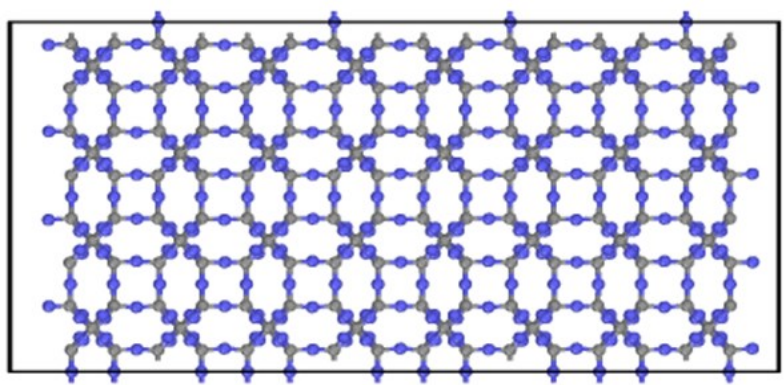

Figure S2. Configuration of the ZIF-8 slab used to perform the CG-MD simulations of the ZIF-8/PVDF slab system. It was obtained by coarsening an AA configuration coming from a previous work.[7]

4) AA-MD simulation for the ZIF-8/PVDF slab system

The simulation made for the AA ZIF-8/PVDF slab system follows the same protocol as the one deployed in many previous works in which MOF/polymer interfaces were studied in AA resolution.[5,7,12-15] The initial configuration was created by putting 21 PVDF chains along the direction normal to the surface of the AA version of the ZIF-8 slab borrowed from previous works.[7] The force field is given by bonded and non-bonded potentials. The latter consist of electrostatic interactions and 12-6 LJ potentials, while the former correspond to bond, 3-body angle, dihedral and improper potentials. The parameters for the potentials concerning interactions between ZIF-8 atoms come from a version of the Zheng et al, 2011 potential for ZIF-8, modified for ZIF-8 surfaces.[7,16] The parameters for the potentials concerning PVDF atoms are the same as the ones used in the AA-MD simulations for the bulk polymer, discussed in section 1. The parameters for the cross 12-6 LJ interactions were determined using the Lorentz-Berthelot mixing rules, while the cross Coulombic interactions were computed taking the charges defined for the intra-phase interactions previously mentioned into account. The timestep for the MD was fixed to 1 fs.

The AA-MD simulation starts with a version of the protocol mentioned in section 1 adapted to study MOF/polymer slab systems in previous works.[5,7,12-15] The modified protocol consists of a sequence of MD steps in which simulations are made at the $NVT_{high}$, NVT and $NP_zT$ ensembles in an alternating manner. The goal is the same as the one of the original protocol, mentioned in section 1 of the SI, with the difference that now the polymer equilibrates in the presence of the MOF. Only the dynamics of the polymer is considered in the protocol, and the pressure in the $NP_zT$ runs concerns only the component normal to the surface of the slab (that is, the projection in the z direction). Aiming to counter the fact that the conformational space of the polymer in the context of the binary system cannot be ergodically sampled in a computationally affordable fashion, 10 independent AA-MD simulations were carried out starting from the microstate attained by the end of the protocol. Similarly to the approach used for the bulk polymer, each of these 10 independent simulations include a set of 10 heating and cooling cycles made in the NVT ensemble with different temperatures and durations that is expected to ultimately lead to configurations belonging to different regions of the conformation space. Subsequent to the 10 heating and cooling cycles associated with each independent simulation, dynamics was performed using the NPT equations of motion. Averaging the results of 10 independent simulations has been proven to yield the same results as those coming from an enhanced sampling simulation, in which the phase space is correctly sampled.[7] The target temperature and pressure were set to 300 K and 1 atm, respectively. The three dimensions of the simulation domain were allowed to change during the dynamics. Exceptionally, the damping constant for the barostat was set to 500 fs. Configurations stemming from each of the independent AA-MD simulations were collected after fluctuations in instantaneous values of potential energy and volume accused the system to be equilibrated. These configurations were used as reference for fitting FM potentials for a CG force field for the binary system.

5) Force matching

Force matching is a potential fitting strategy that allows reaching a CG force field that most accurately reproduces the expected forces (based on a reference) that the beads should experience in a given set of CG configurations. The force matching algorithm starts with defining the form of the force field, usually given by a sum of bonded and non-bonded potentials. The algorithm can accommodate for both analytical and non-analytical potential forms. A set of reference configurations in the CG resolution of choice as well as the forces that the beads should be experiencing in each of them needs to be provided as input. Given the defined form of the potentials, an expression for the force that the CG force field predicts each bead should be experiencing in each reference configuration can be written. The algorithm thus aims to find potential parameters such that, through a conjoint effect of all potentials, the CG force field predicts forces that match the reference values provided in the overall picture (i.e., considering all beads and all reference configurations). This means minimizing the functional shown in equation S1, where **F** denotes the force, and the subscripts *ref*, *CG force field*, *l* and *i* denote the reference force, the force predicted by the CG force field, a given CG configuration and a specific bead index, respectively. In this equation, the total number of beads in the simulation domain and of configurations considered in the FM algorithm are symbolized by N and M, respectively.

$$\chi = \sum_{l=1}^{M} \sum_{i=1}^{N} |F_{ref} - F_{CG\,force\,field}|^2 \tag{S1}$$

In the algorithm, the reference CG configurations and forces are usually obtained from a classical AA-MD simulation of the system. In this case, a set of AA configurations for the equilibrated system are coarsened according to the resolution of choice and the reference forces are computed as a function of the forces experienced by the atoms, which are given by the force field used in the AA simulation. Although the way to compute forces at the CG level as a function of atomic forces is not unique, insights on suitable ways to do so have been provided in previous works.[17,18] Figure S3 shows a simple schematic for the algorithm considering a system of three beads and a CG force field given by a sum of pairwise non-bonded potentials. The equations shown in figure S3 have in the left hand side the reference force the given bead should be experiencing while in the right hand side the forces are those predicted by the CG force field. Given that the pair distances are known from the configurations, the unknowns are the parameters corresponding to the form of the potentials. Since force matching strives to reproduce the reference forces, the potential parameters optimized by the algorithm are such that the equations in figure S3 are satisfied for the overall set of CG configurations used as input. In general, the force matching potentials have non-analytical form, given by piecewise functions such as cubic splines.[19-21] In this case, the parameters are, amongst others, the values of the forces themselves at the pair distances marking the limits of the bin where a given cubic polynomial holds. A very good discussion on what are the parameters of cubic spline potentials can be found elsewhere.[19]

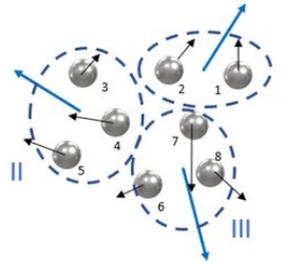

Assuming pairwise non-bonded interactions:
$$F_I = F_{I,II}(parameters) + F_{I,III}(parameters)$$
$$F_{II} = F_{II,I}(parameters) + F_{II,III}(parameters)$$
$$F_{III} = F_{III,I}(parameters) + F_{III,II}(parameters)$$
There will be a set of three equations at each coarsened AA microstate!

Figure S3. Scheme of an AA configuration of 8 atoms (numbered gray spheres) superimposed with its coarsened counterpart according to a mapping that features 3 beads (blue dashed ellipsoids, indexed in roman numerals). The forces experienced by the atoms and corresponding reference forces experienced by the beads are shown using black and blue arrows, respectively.

The cubic spline potentials that are commonly used within FM are also adopted in this work. Notably, as discussed in previous work,[22] the pair distances between ZIF-8 beads in the resolution given in figure 1(a) are not sampled throughout a single continuous interval at ambient conditions in the bulk system, meaning that the corresponding values of forces stemming from the pair potentials cannot be determined in a whole continuous range of pair distances. The same is observed here for the ZIF-8 phase in the binary system. In principle, this is a problem in the context of cubic spline potentials since the parameters are the values of the forces along different points of the domain. Yet, this does not pose an issue, since (1) the FM algorithm implementation in BOCS,[23] software used in this work, allows to successfully carry out the algorithm in several split ranges for which there is sampling of pair distances for a given potential within the CG configurations used as input, and (2) only the (ZIF-8)-(gas) potentials, for which sampling across a single continuous interval of pair distances is always observed at ambient conditions, were further considered.

6) CG force field details

The bonded and non-bonded parameters for the MARTINI potentials used in the hybrid MARTINI/FM force field for ZIF-8/PVDF are summarized in tables S1 and S2, respectively. In the tables, the pair or sequence of three bead types the given potential acts on is informed in accordance with the bead type indexing defined in figures 1(a) and 1(b) of the main text. The potentials concerning interactions between ZIF-8 beads come from model mBM3III, developed in a previous work.[24] Two distinct potentials for angles 212 are defined in this model, whose corresponding values of $K_\theta$ and $\theta_o$ are informed in order in table S1. The values of $K_\theta$ of the angle potentials and $\varepsilon$ of the 12-6 LJ potentials are in kcal/mol, $K_b$ of the bond potentials in kcal/(mol·Å²), $r_o$ and $\sigma$ in Å and $\theta_o$ in degrees. The LJ parameters concern the potential form shown in equation S2.

| ZIF-8 | PVDF |
|---|---|
| Bond 1-2: $K_b$ = 48.3, $r_o$ = 3.02 | Bond 1-1: $K_b$ = 5.5, $r_o$ = 4.82 |
| Angle 121: $K_\theta$ = 1000.0, $\theta_o$ = 169.2 | Bond 1-4: $K_b$ = 19.5, $r_o$ = 3.75 |
| Angle 212: $K_\theta$ = 19.5 & 40.5, $\theta_o$ = 92.7 & 117.9 | Bond 1-2: $K_b$ = 34.0, $r_o$ = 3.97 |
| | Bond 1-3: $K_b$ = 30.2, $r_o$ = 3.99 |
| | Bond 4-3: $K_b$ = 68.0, $r_o$ = 2.81 |

Table S1. Parameters of the bonded potentials featured in the hybrid MARTINI/FM force field used in this work. The form of these potentials can be found in the original publication as well as in the main text.[25,26]

| ZIF-8 | PVDF |
|---|---|
| $(1)_{\text{ZIF-8}}$-$(1)_{\text{ZIF-8}}$ : $\varepsilon = 1.0246$, $\sigma = 4.1$ | $(1)_{\text{PVDF}}$-$(1)_{\text{PVDF}}$ : $\varepsilon = 0.8819$, $\sigma = 4.7$ |
| $(1)_{\text{ZIF-8}}$-$(2)_{\text{ZIF-8}}$ : $\varepsilon = 0.9124$, $\sigma = 4.1$ | $(1)_{\text{PVDF}}$-$(2)_{\text{PVDF}}$ , $(1)_{\text{PVDF}}$-$(3)_{\text{PVDF}}$, $(1)_{\text{PVDF}}$-$(4)_{\text{PVDF}}$: $\varepsilon = 0.8078$, $\sigma = 4.3$ |
| $(2)_{\text{ZIF-8}}$-$(2)_{\text{ZIF-8}}$ : $\varepsilon = 0.8479$, $\sigma = 4.1$ | $(2)_{\text{PVDF}}$-$(2)_{\text{PVDF}}$ , $(2)_{\text{PVDF}}$-$(3)_{\text{PVDF}}$ , $(2)_{\text{PVDF}}$-$(4)_{\text{PVDF}}$ , $(3)_{\text{PVDF}}$-$(3)_{\text{PVDF}}$ , $(3)_{\text{PVDF}}$-$(4)_{\text{PVDF}}$ , $(4)_{\text{PVDF}}$-$(4)_{\text{PVDF}}$ : $\varepsilon = 0.6788$, $\sigma = 4.1$ |

Table S2. Parameters of the non-bonded MARTINI potentials featured in the hybrid MARTINI/FM force field used in this work. These correspond to 12-6 LJ potentials. As several potentials associated to different combinations of PVDF bead pairs are the same, they are presented in the same line.

$$U(r) = 4\varepsilon \left( \left(\frac{\sigma}{r}\right)^{12} - \left(\frac{\sigma}{r}\right)^{6} \right) \quad (S2)$$

A plot of the FM potentials used to model the cross interactions in the MARTINI/FM force field is shown in figure S4. The corresponding potential tables can also be found elsewhere.[27] As discussed in the main text, the values of potential energy at very short pair distances were obtained by fitting an exponential decay function, and a smoothening where original FM data meets data from the function was made. Interestingly, it is possible to see that the FM potentials have a form very similar to the one of 12-6 LJ potentials.

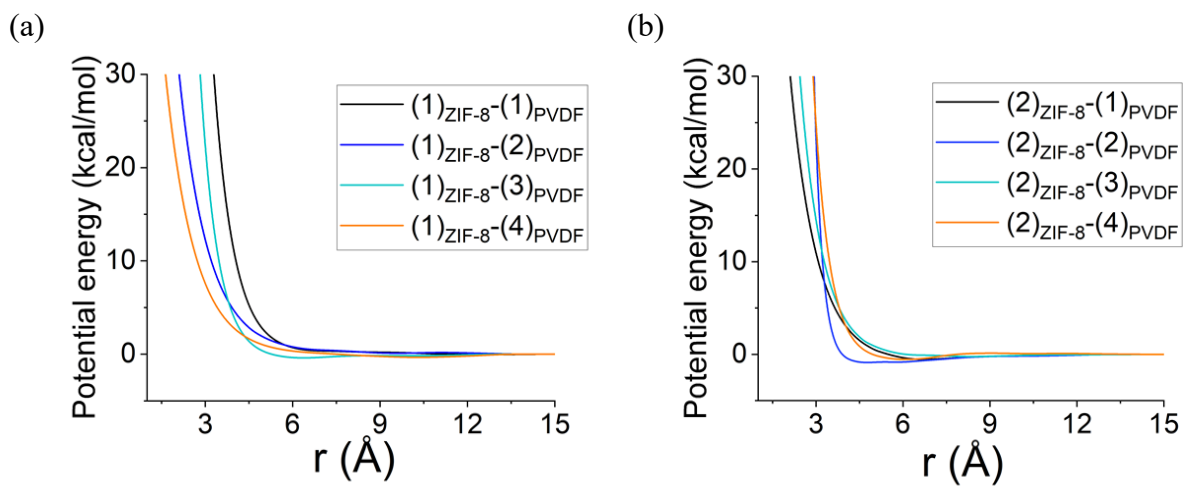

Figure S4. Plot of the cross non-bonded potentials between PVDF and ZIF-8 beads. The caption in the figures indicates the bead type pair a given non-bonded potential acts on. The indexing follows the definition given in figure 1(a) and (b) in the main text.

7) Scheme for nanoparticle surface termination

In this work, three different ZIF-8 nanoparticles are considered to investigate the influence of filler morphology and size on polymer structuration and gas adsorption. Three different nanoparticles were considered: a cubic of radius 69 Å (CB) and two rhombic dodecahedrons of radius 62 Å (RD) and 28 Å (SRD) were considered. As discussed in the main text, the surface of all three nanoparticles investigated in this work are formed by 4- and 6-MR windows. The beads type 1 present at the surface were terminated stoichiometrically and in alternating fashion, so that each undercoordinated Zn bead forming the window is ultimately surrounded by two others that are bonded to four ligands, including a surface (undercoordinated) ligand. The scheme for all the 4- and 6-MR windows is the same and are shown in figure S5, individually. The fully coordinated Zn beads composing the windows are highlighted in red together with the surface ligand bonded to them. Naturally, the scheme for surface 4- and 6-MR that share a Zn is made so that the scheme shown in figure S5 is respected for the two windows individually. Since all surface 4- and 6-MR windows obey the pattern depicted in figure S5, the symmetry inherent to the geometry of the nanoparticle morphology concerning faces, edges and vertices is respected, as mentioned in the main text.

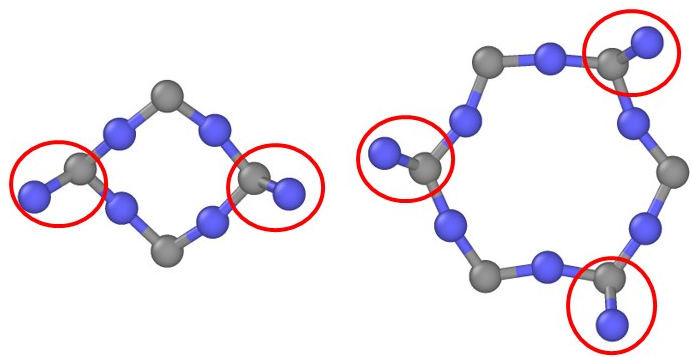

Figure S5. Scheme of termination of the 4- and 6-MR windows found at the surface of the CB, RD and SRD nanoparticles. The red circles highlight surface (i.e., undercoordinated) beads type 2 and the beads type 1 bonded to them. These beads type 1 have four bonds (one of them, lying in the inner portion of the structure, is not shown in the scheme), while the other beads type 1 forming the 4- and 6-MR windows (i.e., those not highlighted) are undercoordinated, having three bonds each (one of their bonds is, again, not shown in the scheme as it is in the inner portion of the structure).

8) CG-MD simulation for ZIF-8/PVDF at the nanoparticle level and methodology for computing density profiles, PVDF ADFs and (ZIF-8)-(PVDF) RDFs

As discussed in the main text, three different ZIF-8/PVDF systems were assembled, each comprising one of the ZIF-8 nanoparticles considered herein in the simulation domain. These are referred to as CB, RD and SRD ZIF-8/PVDF systems throughout the text. An initial configuration for each of these three systems was made as discussed in the manuscript. A protocol to get rid of artificial empty spaces generated when creating the initial configuration was deployed. Aiming for consistency, the protocol used prior to the equilibration and production in the CG-MD simulations for ZIF-8/PVDF at the nanoparticle level is very similar to the one used for the CG ZIF-8/PVDF slab system. At a first moment, the simulation box was steadily deformed while the dynamics of ZIF-8 and PVDF beads was carried out using the NVT equations of motion with the target temperature set to 300 K. The final dimensions of the simulation box were set to be the same in all three directions (cubic simulation box domain) and were calculated based on the estimated volume the corresponding mass of PVDF should occupy in order to meet the experimentally reported density value for the bulk polymer assuming no polymer penetration occurs.[8] Once the final dimensions were reached, the dynamics was carried out using the NPT equations of motion with the target pressure and temperature set to 1 atm and 300 K, respectively, for a total of 1M timesteps. At this stage it is expected that all the artificial empty space existing in the initial configurations has disappeared. A 10 fs timestep was considered for the dynamics aiming to prevent possible numerical instability issues during the simulation. Subsequently, the timestep was increased to 20 fs to perform the equilibration and production. Configurations were then collected in the ranges 66M-78M, 65M-81M and 19M-40M timesteps for the CB, RD and SRD ZIF-8/PVDF systems, respectively.

As mentioned in the main text, instantaneous properties in the SRD ZIF-8/PVDF system started fluctuating around a mean value faster than for the other two systems. Despite the soft variation of mean values observed for the CB and RD ZIF-8/PVDF systems in the ranges in which configurations were collected, these systems can be at least fairly compared to one another as the simulation time considered for both is very alike. Figure S6 shows the fluctuations of instantaneous values of potential energy and volume for the three ZIF-8/PVDF nanoparticle systems within the range of timesteps from which configurations were collected. Finally, figure

S7 shows a snapshot of the CB, RD and SRD ZIF-8/PVDF systems to illustrate how the simulation domain looks.

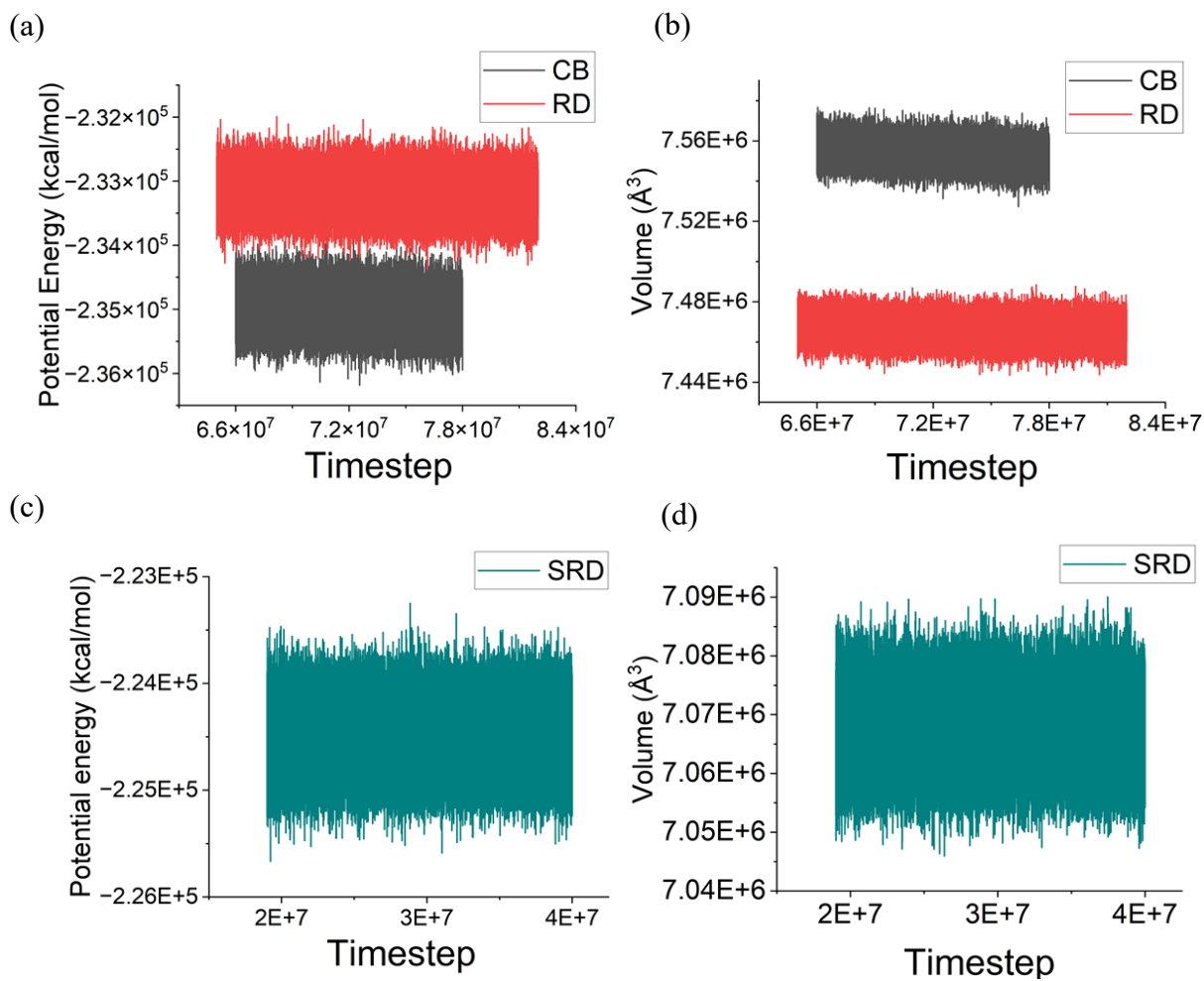

Figure S6. Fluctuations of instantaneous values of (a) potential energy and (b) volume for the CB and RD ZIF-8/PVDF systems in the range of timesteps from which configurations were collected for studying the systems. The instantaneous fluctuations in potential energy and volume values are shown in figures (c) and (d), respectively, for the SRD ZIF-8/PVDF system.

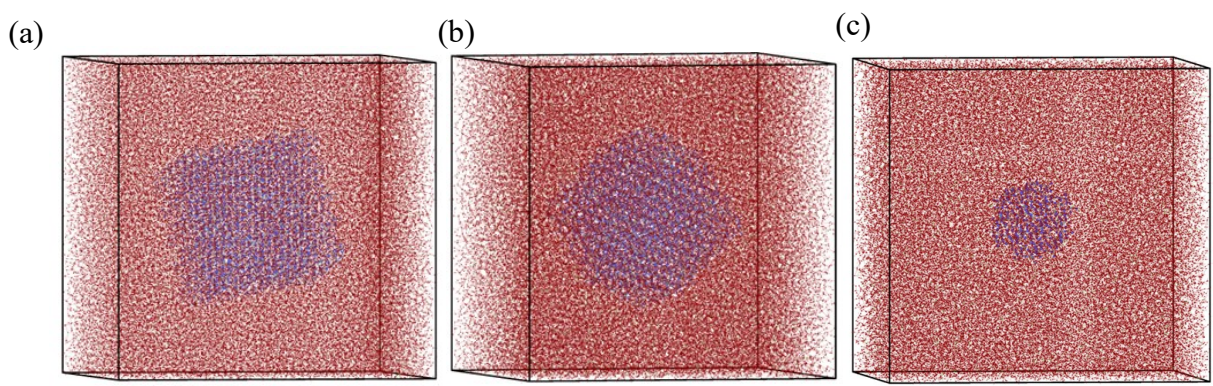

Figure S7. Snapshot of a configuration for the (a) CB, (b) RD and (c) SRD ZIF-8/PVDF system. The snapshots allow seeing the one nanoparticle (beads type 1 and 2 in gray and blue, respectively) composing the simulation domain in each case as well as the polymer surrounding it (all the polymer beads are shown in dark red). All images share the same scale for clarity purposes, even though the volume of the simulation cells is not the same.

The configurations collected were used to compute: (i) the bead density profiles in the simulation domain as well as near/at the nanoparticle surface, (ii) 111 ADFs (bead type indexing in accordance with figure 1(b)) and (iii) (ZIF-8)-(PVDF) RDFs for end-chains PVDF beads for assessing the ZIF-8/PVDF nanoparticle systems as mentioned in the main text. The bead density profiles spanning through the entire simulation domain as well as at the surface of the nanoparticle were computed using in-house python codes, which can be found elsewhere.[27]

To compute the global density profiles, the simulation domain was discretized with ≈0.5 Å thick spherical shells centered in the center of mass of the ZIF-8 nanoparticle and a histogram of bead count in the radial direction was made from several configurations. The histogram was further averaged within all the configurations considered as well as normalized by the volume of the spherical shell corresponding to each bin. The bead density profiles for ZIF-8 and PVDF were built without differentiating bead types within each phase.

For the PVDF bead density profiles at the surface level, the space was discretized in spherical shells 0.25 Å thick centered around the points of assessment of the vertices, edges and faces of the nanoparticle at each of the three systems, shown in figure 3 and discussed in the main text. These vertices, edges and faces are sometimes referred to as different local environments throughout the text. The count of PVDF beads falling within < 12 Å from the point of assessment were considered for the CB and RD ZIF-8/PVDF system. Exceptionally, the threshold was narrowed to < 5 Å for the SRD ZIF-8/PVDF system to avoid overlapping of local environments within such a small sized nanoparticle. Together with information on the distance from the point of assessment (d), the angle, $\alpha$, formed between the PVDF bead, the point of assessment and the center of the nanoparticle was also computed. Each combination of values (d, $\alpha$) corresponds to a circumference in the 3D space, as shown in figure S8. In this figure, the orange, green and blue circles play the role of a PVDF bead, the point of assessment and the center of the MOF nanoparticle, respectively, while the dashed circumference is the region of space where (d, $\alpha$) values are the same. The total count associated to each (d, $\alpha$) combination was then divided by the length of the circumference, yielding linear PVDF bead density profiles, which were then averaged within the total number of configurations considered. It is important to highlight that the circumference region in which the count is performed, shown in figure S8, is not symmetrical, in the sense that not all of its points are surrounded by the same arrangement of ZIF-8 beads. This may be relevant for analyzing the final averaged linear density associated to each (d, $\alpha$) value. Finally, as discussed in the main text, the PVDF bead density profiles concerning local environments that are equivalent to one another within the symmetry of the polyhedra underlying the nanoparticle morphology were averaged.

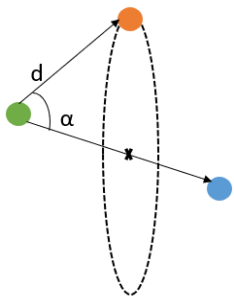

Figure S8. Illustration of the circumference (dashed line, with perspective added) corresponding to the set of points associated with a same combination of values (d, $\alpha$), given the definition of d and $\alpha$. The parameters d and $\alpha$ are also indicated.

Aiming to further assist on the analysis, the values (d, $\alpha$) associated to the center of the 4- and 6-MR windows featured in the different environments were also computed and are presented in table S3. These correspond to the 4- and 6-MR windows whose center sits the closest to the point of assessment of each environment. These values considered the center of the rings as the center of mass of Zn beads composing the given window, and are ultimately the result of an average over all configurations for each ZIF-8/PVDF nanoparticle system.

|    |             | 4-MR                       | 6-MR                       |
|----|-------------|----------------------------|----------------------------|
| CB | Vertices    | d = 8.45 ; $\alpha$ = 54.42   | d = 2.49 ; $\alpha$ = 6.46    |
|    | Edges       | d = 6.29 ; $\alpha$ = 71.05   | d = 4.72 ; $\alpha$ = 64.34   |
|    | Faces       | d = 2.76 ; $\alpha$ = 170.58  | d = 6.18 ; $\alpha$ = 76.23   |
| RD | Vertices g1 | d = 2.34 ; $\alpha$ = 6.86    | d = 8.86 ; $\alpha$ = 42.57   |
|    | Vertices g2 | d = 8.49 ; $\alpha$ = 54.45   | d = 2.60 ; $\alpha$ = 9.58    |

|  | Edges | d = 4.46 ; α = 51.95 | d = 4.05 ; α = 68.27 |
|---|---|---|---|
|  | Faces | d = 6.35 ; α = 71.09 | d = 4.73 ; α = 64.19 |
| SRD | Vertices g1 | d = 2.31 ; α = 6.41 | d = 8.84 ; α = 42.66 |
|  | Vertices g2 | d = 8.49 ; α = 54.45 | d = 2.53 ; α = 5.73 |
|  | Edges | d = 4.44 ; α = 52.19 | d = 4.04 ; α = 68.11 |
|  | Faces | d = 6.33 ; α = 71.32 | d = 4.70 ; α = 64.48 |

Table S3. Average distances in Å, d, between the 4- and 6-MR windows featured in the environments of the different nanoparticles and the corresponding point of assessment. The average angle, α, formed between the center of the window, the point of assessment and the center of the nanoparticle is also shown in degrees.

The 111 ADFs were calculated using an in-house python code. For computing them, the space was discretized in ≈5 Å thick spherical shells centered in the center of mass of the nanoparticle. A 3-body angle value was assigned to a spherical shell when the central atom lay within its domain. The spherical shells span the entire simulation box domain and, for each of them, one ADF for each sequence of three bead types 1 was computed. Each ADF was normalized so that the sum of the function values over all bins were equal to one. This should allow for a fairer comparison between ADFs of different spherical shells as their intrinsic difference in volume would imply a difference in total count. Finally, the RDFs for (PVDF)-(ZIF-8) bead pairs, on the other hand, were computed globally, i.e., without any regard of the positioning of the pair in the overall simulation domain. The calculation was made by LAMMPS during the simulation and output afterwards.

## 9) Benchmark density and RDF data coming from atomistic simulations

Configurations saved during the AA-MD ZIF-8/PVDF simulations described in section 4 were used not only for deriving the (ZIF-8)-(PVDF) FM potentials but also to derive benchmark data to compare the results coming from the CG models. As mentioned in the main text, the structure of the AA slab system was assessed in terms of atomic density profiles, built along the direction normal to the surface of the ZIF-8 slab, and (ZIF-8)-(PVDF) RDFs for the CG resolution. Density profiles were calculated using an in-house python code that discretizes the direction normal to the surface of the slab in bins ≈0.5 Å thick to make the atom count and subsequently divides the count at each bin by the corresponding volume of the region, and outputs the profiles averaged over many AA configurations used. The atomic density profiles were made for the ZIF-8 and PVDF phase individually, with no distinction of atom types within each phase. The RDFs were computed using the coarsened configurations (coarsening made using an in-house python code), which were then used as input in a LAMMPS script to compute RDFs.[28]

As mentioned in the main text, the density of bulk PVDF was also computed in the AA resolution to evaluate if the polymer density in the AA ZIF-8/PVDF slab system matches the bulk one outside the region of overlap with the MOF. This was done considering configurations attained post equilibration in the simulations described in section 1, which concern a thermodynamic state marked by (T = 300 K, P = 1 atm). A value of 0.085 atoms/Å$^3$ was found.

## 10) CG-MD simulations for the ZIF-8/PVDF/CO$_2$ systems

As discussed in the main text, hybrid CG MC/MD simulations were performed for each ZIF-8/PVDF nanoparticle system using a configuration collected within the range considered for studying the empty gas systems (mentioned in section 8) to investigate them under $CO_2$ loading. Specifically for the CB and RD ZIF-8/PVDF systems, the configuration attained at the 78 M timesteps mark of the CG-MD simulation for the binary system was considered. In the MC/MD simulations made for studying gas adsorption, the dynamics of all beads were performed via the NVT equations of motion with the target temperature set to 300 K. Every 10 timesteps, 2 MC moves were attempted. These consisted of $CO_2$ insertion and deletion attempts. The chemical potential of the gas was computed by LAMMPS considering $CO_2$ as an ideal gas (fugacity coefficient = 1) and a gas reservoir at T = 300 K and p = 1 atm.[29] A minimum distance of 1.2 Å from any bead composing the system was set for $CO_2$ additions, aiming to counter possible numerical instability issues due to the molecules being added too close from other beads in the system. For similar reasons, the timestep was set to 1 fs, as advised elsewhere.[29] Two different hybrid simulations starting from the same initial configuration and differing in the random seed number for the MC algorithm were performed for each ZIF-8/PVDF nanoparticle system. These simulations were continued until the amount of $CO_2$ in the system fluctuated around a stable mean value, suggesting that the equilibrium amount of gas at the given (T,P) condition had been reached. The equilibrium amount of gas accused in the two simulations were averaged to get a better estimation.

Aiming to study the gas and polymer structuration in each ZIF-8/PVDF/$CO_2$ system, the configuration attained by the end of one of the two hybrid MC/MD simulations was used to further perform an MD simulation. The NPT equations of motion were considered for the dynamics, with target pressure and temperature set to 1 atm and 300 K. As no more gas insertions or deletions were made, the timestep was raised to 10 fs, a value still inferior to the one used for the binary system to accommodate for the fact that the system is denser due to the gas adsorption. Once the $CO_2$ loaded systems were equilibrated, configurations were collected to compute bead density profiles mentioned in the main text.

11) Results: Bead density profiles for the ZIF-8/PVDF nanoparticle systems

Figures S9, S10 and S11 show the PVDF bead density profiles around the points of assessment of the different local environments in the CB, RD and SRD ZIF-8/PVDF nanoparticle systems that are not shown in the main text. The x and y axis in the figures show the values of $\alpha$ and $d$ associated to each point in space, while the color scale indicates the magnitude of the linear density (in bead/Å units) associated with each given combination $(d,\alpha)$. The figures also show the combination of values $(d,\alpha)$ associated with the center of the 4- and 6-MR windows featured in the different environments, and shown in table S3, to facilitate the analysis.

It is possible to spot non-zero values close to the 6-MR window in all environments of the three ZIF-8/PVDF nanoparticle systems. Upon identifying combinations $(d,\alpha)$ that correspond to regions inside and outside the nanoparticle domain with the assistance of figure 3, it is possible to see that the non-zero polymer density values form a path that goes through the 6-MR window, evidencing thus polymer penetration in all cases. On the other hand, most of the plots accuse no presence of polymer at points $(d,\alpha)$ close to the center of the 4-MR window, suggesting that no polymer penetration occurs through these windows in these cases. For some local environments, there is non-zero polymer density near the 4-MR window, but this could be ascribed to a count at a region of space far from its center but that shares the given value of $(d,\alpha)$ within the circumference the combination corresponds to.

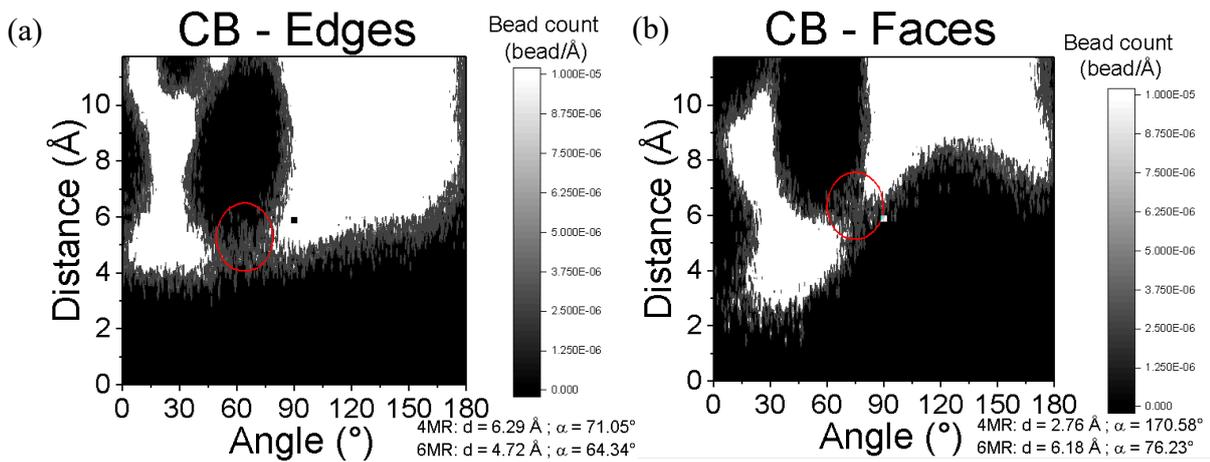

Figure S9. Density profiles for the (a) edges and (b) faces of the CB ZIF-8/PVDF system. The color scale indicates the density value at a given distance (d) from the point of assessment and angle ($\alpha$) formed between the PVDF bead, the point of assessment and the center of the nanoparticle. The values of (d,$\alpha$) associated with the center of the 4- and 6-MR windows nearest to the point of assessment of the given window are shown below to facilitate the analysis.

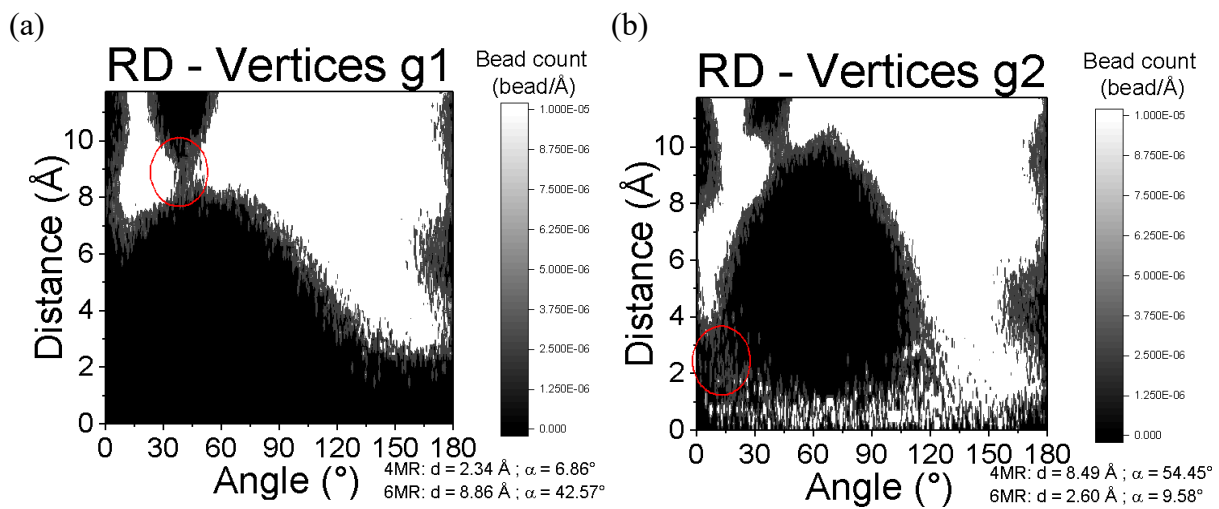

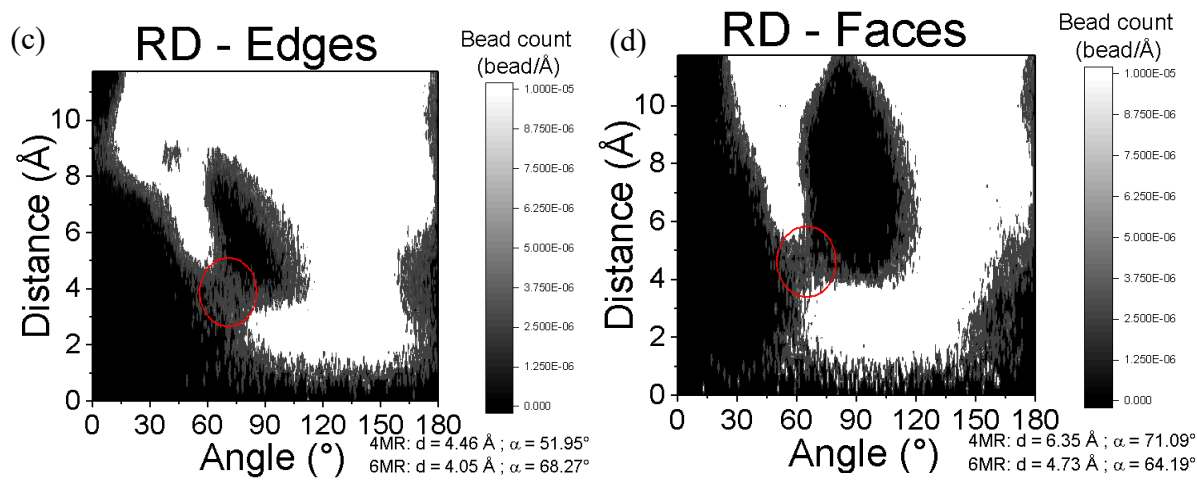

Figure S10. Density profiles for the (a) vertices g1, (b) vertices g2, (c) edges and (b) faces of the RD ZIF-8/PVDF system. The color scale indicates the density value at a given distance (d) from the point of assessment and angle ($\alpha$) formed between the PVDF bead, the point of assessment and the center of the nanoparticle. The values of (d,$\alpha$) associated with the center of the 4- and 6-MR windows nearest to the point of assessment of the given window are shown below to facilitate the analysis.

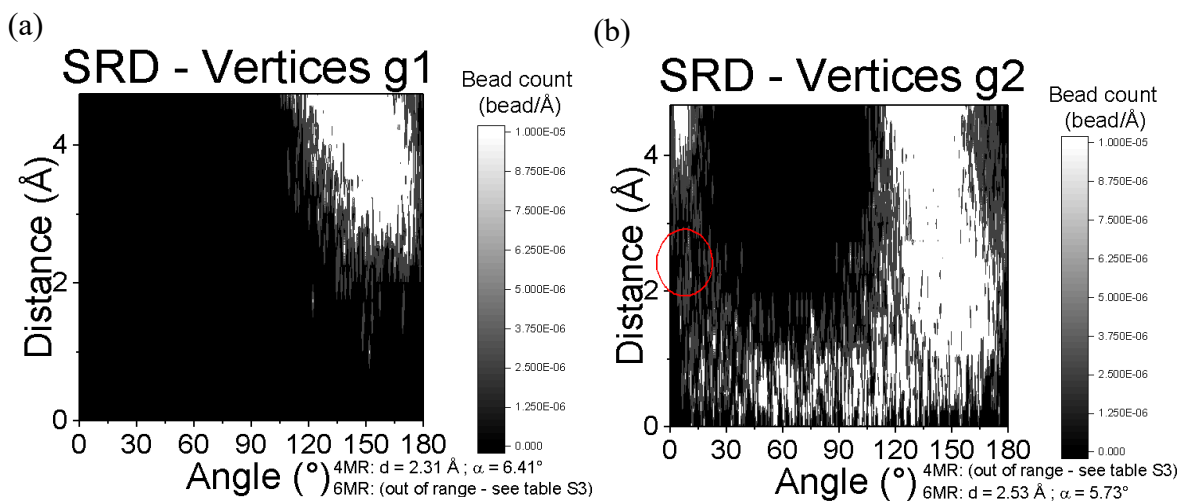

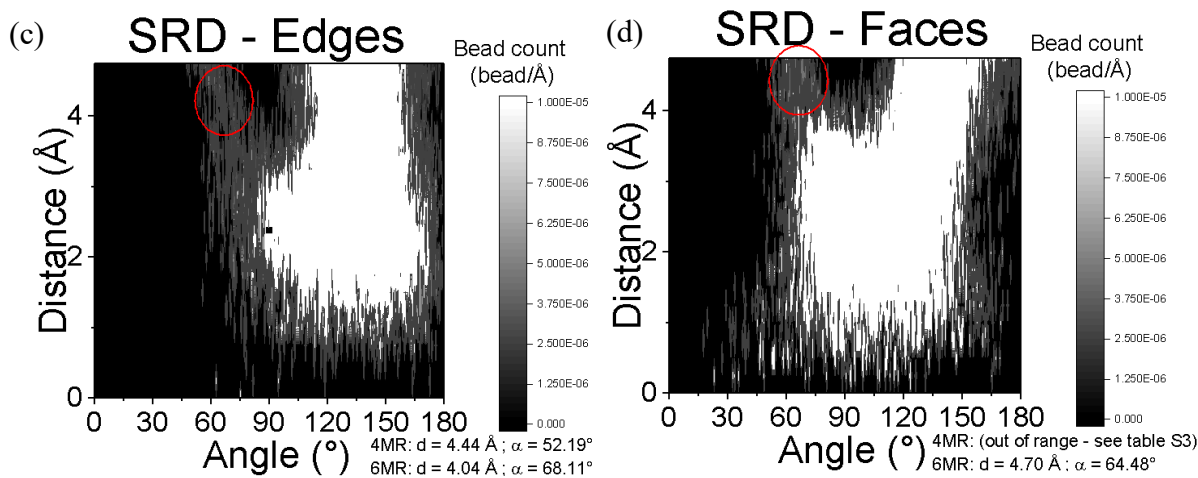

Figure S11. Density profiles for the (a) vertices g1, (b) vertices g2, (c) edges and (b) faces of the SRD ZIF-8/PVDF system. The color scale indicates the density value at a given distance (d) from the point of assessment and angle (α) formed between the PVDF bead, the point of assessment and the center of the nanoparticle. The values of (d,α) associated with the center of the 4- and 6-MR windows nearest to the point of assessment of the given window are shown below to facilitate the analysis.

12) Results: $CO_2$ bead density profiles for the RD and SRD ZIF-8/PVDF/$CO_2$ systems

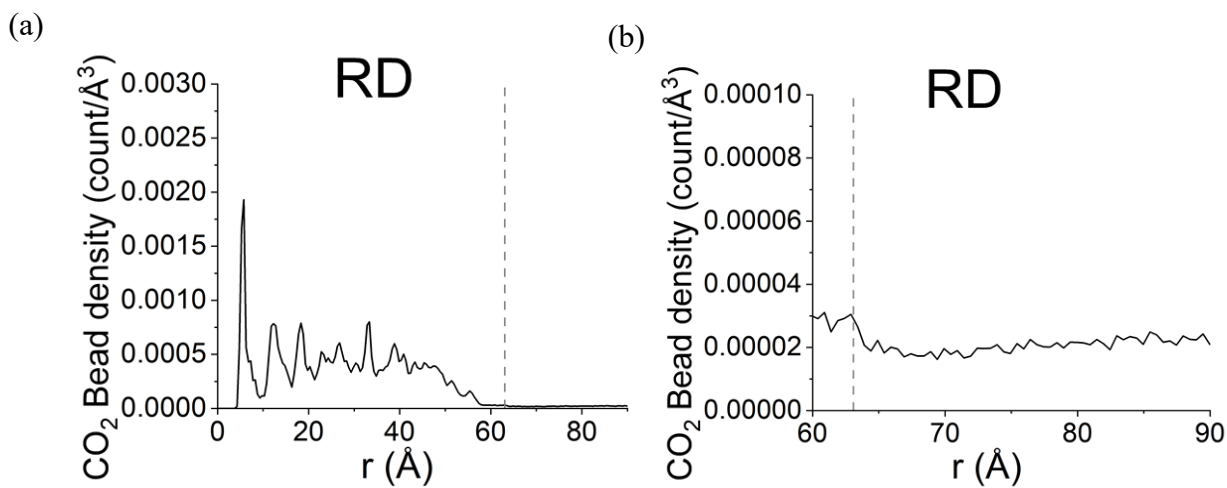

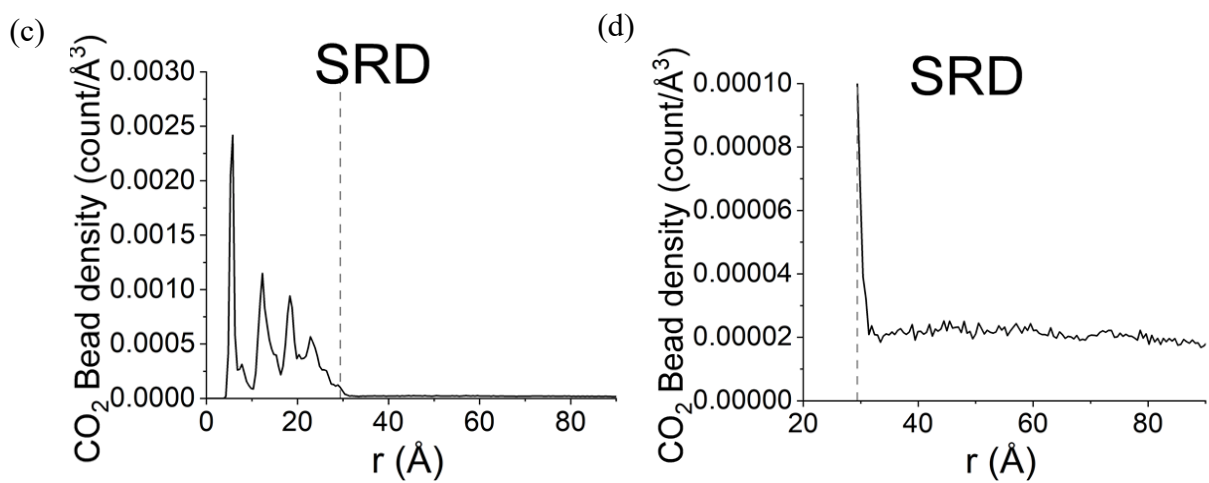

Figure S12. CO$_2$ bead density profile along the simulation domain and at and near the PVDF phase outside the MOF/polymer overlap region for the (a)-(b) RD and (c)-(d) SRD ZIF-8/PVDF systems. The dashed lines mark the largest value of r at which a non-zero ZIF-8 bead count occurs, serving thus to delimit the ZIF-8 nanoparticle phase within the simulation domain.